%% file: main.tex
\documentclass[11pt]{article}
\usepackage[a4paper, margin = 1.2in]{geometry}
\usepackage{amsfonts,amsmath,amssymb} % Mathematical symbols.
\usepackage{booktabs} % For publication quality tables for LaTeX
\usepackage{graphicx}
\usepackage{xcolor}
\usepackage{listings} % For source code snippets.
\usepackage{xspace}
\usepackage{cite}
\usepackage{url}
\usepackage{breakurl}
\usepackage{authblk}
\usepackage{xcolor}

\definecolor{authorA}{rgb}{0.8, 0, 0} % Dark Red
\definecolor{authorB}{rgb}{0, 0, 0.8} % Dark Blue
\definecolor{authorC}{rgb}{0, 0.8, 0} % Dark Green

\title{Cybercrime and Prevention: Colonel
Blotto in Social Engineering}

\author[1]{Gergely Benkő}
\author[2]{Katalin Parti}
\author[1]{Gergely Biczók}
\affil[1]{CrySyS Lab, Budapest University of Technology and Economics\\\texttt{benkogergely@edu.bme.hu, biczok@crysys.hu}}
\affil[2]{Virginia Tech\\\texttt{kparti@vt.edu}}
\date{}
\begin{document}

\maketitle

\begin{abstract}
\input{content/1_abstract}
\end{abstract}

\input{content/2_introduction}
\input{content/3_background}
\input{content/4_basegame}
\input{content/5_extendedgame}
\input{content/6_conclusion}

\bibliographystyle{unsrt}
\bibliography{mybib.bib}

%\printbibliography
%\appendix
%\input{state}

\end{document}

%% file: content/1_abstract.tex
%\biczokg{This is how to use colored text}
Cybercriminals increasingly target the human factor rather than continuously advancing technological defense mechanisms. Consequently, institutions that allocate substantial resources to strengthening their cybersecurity infrastructure may remain vulnerable if a deceived employee voluntarily transmits sensitive information or financial assets to attackers. Therefore, alongside the implementation of technological defense mechanisms, particular emphasis must be placed on mitigating human vulnerabilities, which can be achieved through preventive awareness programs. However, such training activities can only be effective if they are organization- and context-specific.

In this paper, we develop two Colonel Blotto game models to determine the optimal allocation of defensive resources across dominant social engineering attack vectors. We ground the models in Routine Activity Theory (RAT), borrowed from criminology, that describes crime as an event involving a motivated offender, a suitable target, and the absence of a capable guardian. Next, we quantify relevant factors via the VIVA (Value, Inertia, Visibility, Accessibility) framework, and operationalize the models by feeding real-world cybercrime data into them. The first model investigates optimal population-level prevention, focusing on nation-states as defenders; we present and compare use cases of three different countries. The second model focuses on the organization as a decision-maker; here, we analyze five use cases involving organizations of different characteristics. Our results demonstrate that theoretically grounded and data-driven models can provide decision support to policymakers and organizational leaders in allocating their efforts optimally to prevent social engineering attacks and improve their overall cyber resilience. 

%% file: content/2_introduction.tex
\section{Introduction}
\label{sec:introducion}
Over the past decade, malicious activity in cyberspace has increased markedly in both frequency and scale, affecting an increasingly broad range of sectors and organizations. In response, regulators and international economic alliances have implemented comprehensive cybersecurity requirements for all major stakeholders~\cite{nist80053r5upd1}. Consequently, both public institutions and private sector organizations are allocating substantial resources to cybersecurity. A major direction for advanced defense solutions is to leverage artificial intelligence (AI) to detect and mitigate threats at an early stage. As a result, digital systems are becoming more resilient, even if the uptake of secure software development methodologies is lagging, and foundational cybersecurity training has become a standard expectation for IT professionals. Nevertheless, while there is no universally accepted standardized methodology for estimating the aggregate economic impact of cybercrime, all sources point toward the continued rise in reported financial damages attributed to cybercriminal activity~\cite{ic3report2024}.

While there are multiple potential explanations for this trend, including i) the omnipresence of networked IT solutions across all types of organizations, ii) the complexity of software supply chains, and iii) 
the adversarial use of AI for finding zero days, the ever-improving effectiveness of social engineering attacks should not be understated.
In social engineering attacks, adversaries exploit a wide range of psychological manipulation techniques~\cite{hadnagy2010social}, allowing them to bypass even well-engineered technical controls. From the attacker’s perspective, the barrier to entry is low: basic computer skills and familiarity with social media platforms can be sufficient to reach (and potentially deceive) millions of targets at global scale. This favorable cost–benefit profile has contributed to a gradual shift of conventional criminal activity into the digital domain. Moreover, because social engineering tactics are highly adaptable and continue to evolve~\cite{hadnagy2018social} with the emergence of AI, they are also extensively leveraged by state-sponsored actors, supporting objectives ranging from intelligence collection to initial access into protected systems~\cite{mitre2022socialengineering}. While contemporary directives and regulations emphasize the importance of user training, preventive programs are often designed around generic global trends rather than the specific institutional, cultural, and threat-context characteristics of a given organization or nation-state. As a result, such programs may deliver limited effectiveness while imposing avoidable costs on organizations and consuming disproportionate resources on the side of public authorities.

It is evident that tailoring preventive campaigns to the characteristics of concrete organizations and governments and optimizing spending across different social engineering attack vectors are imperative for effective defense. As a response, in this paper, we marry the core principles of cyber victimology, real-world data on the current trends of cybercrime and cyberdefense, and game-theoretical optimal resource allocation, to create a decision support framework for both public policymakers and organizational leaders. Specifically, we develop two Colonel Blotto game models to determine the optimal allocation of defensive resources across dominant social engineering attack vectors. We base our analytical framework on Routine Activity Theory (RAT) from criminology, which conceptualizes crime as the convergence of a motivated offender, a suitable target, and the absence of a capable guardian. To operationalize this perspective for cybercrime, we employ the VIVA framework, assessing Value, Inertia, Visibility, and Accessibility, to quantify key factors that influence victimization risk. Empirical cybercrime and cyberdefense statistics are then used to instantiate the models, bridging theoretical constructs with real-world trends. Our first model examines \emph{optimal population-level prevention} strategies, treating nation-states as primary defenders, and illustrates its application through comparative analyses of three countries with differing profiles. The second model shifts the focus to the \emph{organizational level}, analyzing five organizations with distinct characteristics to evaluate how internal decision-making affects expected loss attributed to social engineering. Collectively, the results indicate that combining theoretically grounded frameworks with data-driven modeling can yield actionable insights for both policymakers and organizational leaders, guiding the strategic allocation of resources to prevent social engineering attacks and, thus, mitigate cybercrime.

The rest of the paper is structured as follows. Section \ref{sec:background} briefly introduces cybercrime and its characteristics, lays the foundations of cyber victimology, and makes the reader acquainted with the Colonel Blotto game. Next, Section~\ref{sec:basegame} constructs our first game focusing on optimal defense at the national level based on country-level statistics and presents use cases on Hungary, Finland, and the United States. Afterwards, Section~\ref{sec:extended} refines the model to capture the heterogeneity of battlefields, making it suitable for organization-level resource allocation. Finally, Section~\ref{sec:conclusion} presents the limitations of our models and potential future work, and concludes the paper.

%% file: content/3_background.tex
\section{Background}
\label{sec:background}

Here, we briefly introduce the necessary background for cybercrime and social engineering, cyber victimology, and the Colonel Blotto game.

\subsection{Cybercrime, social engineering, and economic impact}

INTERPOL defines cybercrime as a type of criminal activity that involves the use of computer networks either as tools for committing crimes or as targets themselves. This definition encompasses both ``pure'' cybercrimes, such as unauthorized access or disruption of data and systems, and traditional offenses that are facilitated by digital technologies~\cite{interpol2022cybercrime}. In contrast, the European Union Agency for Cybersecurity (ENISA) provides a more concise definition, describing cybercrime as any illegal activity that involves the use of a computer device or network~\cite{enisa2017terminology}. Wall et al.~\cite{wall2024cybercrime} categorize cybercrimes into three main types, which reflect the direction, purpose, and technical execution of the criminal act: \textit{Cybercrimes Against the Machine, Cybercrimes Using the Machine, and Cybercrimes In the Machine}. These definitions provide a foundation for understanding the diverse and evolving landscape of cybercrime, which has become a pervasive global threat with profound economic, social, and organizational consequences.

Malicious activity in cyberspace now encompasses a wide spectrum of offenses, ranging from financial fraud, identity theft, and phishing campaigns targeting individuals, to ransomware attacks, supply chain intrusions, and large-scale data breaches affecting corporations and critical infrastructure. For the general public, these attacks can result in direct financial losses, erosion of personal privacy, and long-term psychological stress, while also undermining trust in online platforms and institutions. Large-scale attacks may generate systemic risks, such as disruptions to essential services or cascading losses in digital commerce, which amplify societal vulnerability. Organizations face an even broader and more complex set of consequences: beyond operational disruption and revenue loss, cyber incidents can inflict reputational damage, erode stakeholder confidence, trigger regulatory penalties, and necessitate substantial investments in cybersecurity measures, insurance, and forensic investigation. Many indirect impacts—such as lost productivity, diminished brand value, and market uncertainty—remain difficult to quantify, and variations in reporting, jurisdictional definitions, and disclosure practices further obscure the total economic burden. According to projections by Cybersecurity Ventures, the global cost of cybercrime is measured in trillions of dollars, comparable to the economies of the world’s largest countries, and is growing at an estimated annual rate of 10–15\%~\cite{cyberventures2025cost}. A recent study by the World Bank highlights the complexity and heterogeneity of cybercrime-related costs, noting that indirect damages are often severely underestimated and calling for standardized definitions and unified data collection frameworks to improve transparency and comparability~\cite{worldbank2023cybercosts}. Consistent with these observations, Moore notes that measuring the economic impact of cybercrime remains an open question due to methodological inconsistencies, underreporting, and the diffuse nature of indirect costs~\cite{cybok-security-economics-kg}. These trends emphasize the critical importance of addressing not only technical vulnerabilities but also the social, behavioral, and organizational factors that shape exposure to risk, highlighting the need for frameworks that integrate both human and technological dimensions of defense.

Within this context, social engineering has emerged as one of the most effective and scalable forms of cybercrime, exploiting human cognition, trust relationships, and routine organizational practices to circumvent even robust technical controls. Unlike attacks that rely solely on software vulnerabilities, social engineering manipulates psychological and social levers to induce targeted individuals to reveal sensitive information or perform actions that compromise security. As documented by Hadnagy~\cite{hadnagy2010social, hadnagy2018social}, these attacks employ diverse techniques, including pretexting, phishing, baiting, and elicitation, that exploit human tendencies such as trust, curiosity, fear, and the desire to help. Recent advances in artificial intelligence further amplify the attacker’s capabilities by lowering the cost and increasing the sophistication of deception: large-scale language models can generate highly convincing, context-aware phishing messages; AI-enabled voice synthesis facilitates vishing attacks that impersonate trusted colleagues or executives; and synthetic media technologies, including deepfake video and manipulated images, create novel avenues for fraud, coercion, and reputational harm. The ENISA Threat Landscape 2024 demonstrates that a considerable proportion of detected social engineering incidents are explicitly directed against the general public, and while institutional targets bear the higher overall share of attacks, significant concentrations occur in public administration, banking and finance, and digital infrastructure~\cite{enisa2024}. The effectiveness of social engineering attacks is reinforced by cognitive rationalizations described in the theory of techniques of neutralization, which explain how offenders justify deviant behavior—denying harm, blaming the victim, or appealing to a higher purpose—thus reducing moral and social constraints that would otherwise inhibit criminal activity. Social engineering therefore represents a convergence of low technical barriers, high potential reach, and psychologically reinforced offender motivation, allowing attackers—from opportunistic criminals to state-sponsored actors—to operate efficiently at scale. Hadnagy’s work further highlights that successful attacks exploit situational context, organizational culture, and routine behaviors, demonstrating that vulnerability is shaped as much by human and institutional factors as by technological defenses. Understanding these dynamics provides a natural rationale for structured theoretical frameworks such as \emph{Routine Activity Theory (RAT)}, which formalizes the interaction between motivated offenders, suitable targets, and absent or ineffective guardians, and can be operationalized through quantitative approaches such as the \emph{VIVA framework} to assess risk and guide prevention strategies.

\subsection{Cyber Victimology}
Routine Activity Theory (RAT), developed by Cohen and Felson~\cite{cohen1979social}, is one of the most widely cited criminological frameworks for understanding the circumstances under which crimes occur. RAT posits that criminal acts are more likely when three elements converge in time and space: a motivated offender, a suitable target, and the absence of a capable guardian. Unlike explanations that focus solely on offender characteristics, RAT emphasizes the role of situational and environmental factors, including the behaviors and routines of potential victims, as well as the availability of preventive measures. This perspective allows for the identification of opportunities for crime and offers guidance on how to reduce vulnerability through changes in routines or protective interventions.

As criminal activity has increasingly migrated to digital spaces, researchers have explored the applicability of RAT to online environments. Leukfeldt and Yar argue that the core elements of RAT, asuitable target and the absence of capable guardianship, remain central to understanding cybercrime~\cite{leukfeldt2016applying}. Pratt et al. analyzed the online behavior of adults and found that frequent online activities, such as shopping or active use of social media, increase the likelihood of becoming a target of internet fraud~\cite{pratt2010routine}. Choi~\cite{choi2008computer} similarly finds that online routines and lack of security measures amplify exposure to cybercriminals. Conversely, the presence of effective technological, institutional, and social safeguards can mitigate risk. In cyberspace, suitable targets include not only personal and financial information, but also digital devices, online accounts, and organizational information systems. Capable guardians can take many forms: technological protections such as firewalls, antivirus software, and multi-factor authentication; institutional measures including corporate policies, security protocols, and regulatory compliance; and social guidance, such as peers or family members offering advice and oversight. These protective measures could operate in parallel, creating layers of defense that influence the probability of victimization.

%%%revision new paragraph
In the context of social engineering, however, Routine Activity Theory (RAT) requires a more interactional and organizational refinement. Rather than conceptualizing target suitability solely in terms of exposure or the possession of valuable assets, we argue that suitability in social engineering is produced through a target’s embeddedness in communication flows, trust relations, and routine decision-making processes. In other words, offenders do not merely identify suitable targets; they actively construct suitability through impersonation, urgency, authority cues, and rapport-building. This suggests that capable guardianship should likewise be understood more broadly, extending beyond technical safeguards to include procedural, social, and organizational protections such as callback verification, multiple authorization, escalation protocols, peer consultation, and a workplace culture that legitimizes slowing down and questioning unusual requests. This refinement remains consistent with prior applications of RAT to cybercrime, which have shown that online routines, digital exposure, and guardianship shape victimization risk, while also recognizing that phishing and related forms of social engineering are communicative processes that exploit institutional roles and routine practices rather than only individual vulnerability~\cite{Holt26112008,leukfeldt2014phishing,leukfeldt2016applying,10.1093/bjc/azv011}. Accordingly, in social engineering, crime opportunity should be understood not simply as the convergence of offender, target, and absent guardianship, but as the convergence of a motivated offender with persuasive access to a target within an insufficiently guarded communication environment.
%%%%end revision

The VIVA framework (Value, Inertia, Visibility, and Accessibility) offers a complementary lens to operationalize target characteristics from the perspective of offenders~\cite{cohen2010social}. \emph{Value} represents the potential gain an offender expects from exploiting the target; \emph{Inertia} reflects the effort required to compromise or move the target; \emph{Visibility} captures how easily the target can be observed or discovered; and \emph{Accessibility} measures how readily the offender can reach and exploit the target. While VIVA has often been applied in studies focusing on older adults~\cite{lichtenberg2013psychological, parti2023capable, parti2022elder, parti2025wisdom}, the framework is broadly applicable to all populations and organizational contexts. Each of the four dimensions can be interpreted as a quantitative score, which allows researchers and practitioners to plug in real-world data and systematically assess the relative attractiveness or vulnerability of different targets. For example, high-value data, poorly secured accounts, and frequent online activity combine to make targets more attractive, while strong technical protections and active monitoring reduce exposure. Parti et al.~\cite{parti2025wisdom} compared younger (18–35 years) and older (60+ years) adults as targets for online fraud and found that younger individuals were more often targeted by fast-paced scams appealing to curiosity or quick financial gain, such as phishing or fraudulent job offers, whereas older adults were more likely to fall victim to scams relying on trust-building and longer interactions, such as romance or technical support scams. This comparison illustrates how target characteristics and routines influence offender strategy and demonstrates the general applicability of RAT and VIVA across different demographic groups.

A key aspect of operationalizing RAT in cyberspace is considering attacker motivation and type. For the general population, attackers are predominantly opportunistic cybercriminals seeking financial gain through fraud, phishing, identity theft, or ransomware. In contrast, organizations may face more sophisticated threats, including nation-state actors and Advanced Persistent Threats (APTs), whose motives extend beyond immediate financial gain to include espionage, intellectual property theft, disruption of operations, or geopolitical influence~\cite{alshamrani2019survey}. RAT and VIVA remain applicable in both cases, but the nature of the offender affects how targets are selected, how guardianship is circumvented, and which aspects of VIVA (e.g., value or accessibility) are weighted more heavily in attacker decision-making. Integrating attacker profiles into the analysis allows for a more precise understanding of risk and informs the design of preventive strategies tailored to different threat actors.

%%%%revision new paragraph
This distinction also requires a refinement of how Routine Activity Theory is applied to state-sponsored actors. We do not suggest that nation-state cyber operations are explained by situational opportunity in the same way as purely opportunistic cybercrime. Rather, strategic actors typically select broad targets in line with geopolitical, intelligence, military, or economic objectives, as recent research on state-sponsored cyber espionage makes clear~\cite{10.1177/00223433231214417,10.1093/jogss/ogab028}. However, even when target selection is strategically directed, the concrete execution of cyber operations still depends on opportunity structures at the tactical and operational level, including accessible entry points, exposed personnel, weak organizational routines, and inadequate guardianship. In practice, this helps explain why state-linked and advanced persistent threat campaigns so often rely on spear-phishing, social engineering, and other forms of human-layer exploitation as initial access vectors rather than exclusively on highly bespoke technical intrusions~\cite{10.1093/cybsec/tyad023,10.1093/cybsec/tyaa020}. Accordingly, in our framework, RAT and situational perspectives are not offered as complete explanations of state cyber strategy; instead, they illuminate how strategically chosen operations are translated into practicable attacks through the exploitation of local vulnerabilities and uneven defensive conditions~\cite{10.1093/cybsec/tyv003,10.1093/cybsec/tyaa013}. This narrower conceptualization better captures the coexistence of strategic direction and situational dependence in social engineering against both public institutions and organizations.
%%%end revision

Empirical studies support the utility of RAT and VIVA also in a broader cyber context. Online behaviors serve as predictive factors for victimization, much as physical routines do in traditional criminology. Individuals and organizations with high exposure, insufficient technical safeguards, or limited awareness of social engineering tactics are more likely to be targeted. Social and technical forms of guardianship, ranging from software protections to institutional policies and guidance from peers, significantly influence risk, often mitigating the impact of otherwise attractive target characteristics~\cite{kemp2023consumer, kemp2024worry}. Integrating RAT with VIVA yields a structured, theory-driven approach for analyzing cyber risk. As such, it enables a nuanced understanding of how offender motivation interacts with target properties and protective mechanisms, allowing policymakers and organizational leaders to prioritize interventions and allocate resources efficiently. By formalizing the relationships among offender motivation, target characteristics, and the presence or absence of guardians, this approach allows for systematic identification of vulnerabilities, supports the design of preventive strategies, and provide actionable guidance for individuals, organizations, and policymakers seeking to mitigate social engineering. The ability to quantify VIVA dimensions using real-world data further allows these theoretical frameworks to be operationalized into predictive models, forming a bridge towards data-driven optimization, such as the population- and organization-level game models we construct and analyze later.

\subsection{Colonel Blotto Games in Cyber Defense}

The Colonel Blotto game is a classical model in game theory for studying how strategic actors allocate scarce resources across multiple, simultaneously contested objectives~\cite{borel1921theorie,roberson2006colonel}. In its canonical form, two players each possess a fixed resource budget (e.g., troops, money, time, or effort) and must distribute it across several battlefields. Each battlefield is won by the player who commits more resources to it, while ties are resolved according to a predetermined rule. The players’ objective is to maximize their total payoff, typically defined as the number of battlefields won or a weighted sum when battlefields differ in importance. Because one player’s success necessarily comes at the expense of the other, the standard Colonel Blotto game is naturally modeled as a zero-sum competition.

A defining feature of the Colonel Blotto setting is the interdependence of allocation decisions: assigning additional resources to one battlefield necessarily reduces what remains for others. Moreover, choices are made simultaneously, meaning that each player must act under strategic uncertainty about the opponent’s allocation. This makes predictability costly: a deterministic allocation can be exploited by an opponent who anticipates where resources are concentrated or neglected. Consequently, optimal play often requires randomization, and equilibrium behavior is commonly represented through mixed strategies, i.e., probability distributions over possible allocations. These characteristics make Colonel Blotto games a particularly useful abstraction for cyber defense, where defenders must allocate limited security resources across many assets, vulnerabilities, or attack surfaces, while attackers strategically select targets and adapt their effort in response.

\noindent\textbf{Game-theoretical concepts. }In this framework, the \textit{players} are the strategic decision-makers, most naturally interpreted as attackers and defenders. A player’s \textit{strategy} specifies a complete allocation of its budget across battlefields: a \textit{pure strategy} is a single concrete allocation, whereas a \textit{mixed strategy} randomizes over multiple allocations to reduce predictability. The \textit{utility} (or payoff) function maps the resulting battlefield outcomes into a single measure of performance, such as the number of battlefields won or a weighted sum reflecting heterogeneous importance across targets. Finally, a \textit{Nash equilibrium} is a strategy profile in which no player can improve its payoff by unilaterally deviating, given the other player’s strategy. In many Colonel Blotto variants, pure-strategy Nash equilibria do not exist, while mixed-strategy equilibria do, reflecting the strategic value of unpredictability. However, as the number of battlefields grows and the set of feasible allocations expands, equilibrium computation can become analytically and computationally challenging; in such cases, approximate equilibrium concepts and learning-based methods are often employed. One prominent approach is regret-based learning~\cite{hart2013simple}, in which strategies are iteratively updated in proportion to the advantage that alternative allocations would have provided in hindsight, yielding practical approximations of equilibrium behavior in large strategy spaces.

\noindent{\textbf{Variants of Colonel Blotto.} The classical Colonel Blotto model considers two players distributing limited resources across multiple battlefields, where each battlefield is awarded to the player allocating the greater share of resources and overall success is determined by the number (or value) of battlefields won. Over time, several variants have been introduced to better capture real-world competitive allocation problems, including cybersecurity contexts where actors, targets, and outcomes may be heterogeneous and uncertain. Common dimensions of variation include \textit{symmetric vs.\ asymmetric} games (equal vs.\ unequal total resources), \textit{homogeneous vs.\ heterogeneous} battlefields (equal vs.\ unequal battlefield values), \textit{discrete vs.\ continuous} allocation (indivisible units vs.\ divisible effort), \textit{static vs.\ dynamic} interaction (one-shot simultaneous allocation vs.\ sequential or repeated play), \textit{deterministic vs.\ stochastic} outcomes (winner-takes-all vs.\ probabilistic contest success functions such as lottery Blotto), \textit{complete vs.\ incomplete information} (full knowledge vs.\ uncertainty about resources or valuations), and \textit{multi-player} extensions with more than two competitors. These variants preserve the central trade-off of the model, concentrating resources on high-priority targets versus maintaining broad coverage, while enabling a closer approximation of operational cyber defense decision environments. Note that the foundational theory of Blotto games is still a fertile ground for research.

\noindent{\textbf{Colonel Blotto in cybersecurity applications.} A perennial challenge in cybersecurity is that, unlike many other organizational investments, it rarely generates direct and observable profit. Effective multi-layered defenses tend to remain largely invisible in day-to-day operations, while their costs are immediately apparent in budgets, staffing requirements, and operational overhead. As a result, many organizations converge toward compliance-driven security postures and implement only the legally mandated minimum, unless they have experienced major incidents in the past or operate under sustained geopolitical pressure that elevates cyber risk to a strategic priority. This gap between persistent threat exposure and constrained defensive capacity motivates the use of game-theoretic models that explicitly treat cybersecurity as a resource allocation problem under strategic opposition. Chia and Chuang were among the first to apply Colonel Blotto to cybersecurity, analyzing phishing campaigns as a strategic contest~\cite{chia2011colonel}. Attackers allocate limited effort across multiple potential victims, while defenders distribute monitoring and protective resources to reduce exposure. The study shows that even modest defensive allocations, if optimally distributed, can substantially decrease the success of large-scale phishing campaigns and highlights the value of mixed strategies and probabilistic defense in minimizing predictability. Chia et al. extend this approach to web security, modeling asymmetric ``whack-a-mole'' conflicts in which attackers and defenders continuously reallocate resources across targets in a dynamic, guerrilla-style engagement~\cite{chia2016whack}. Their results emphasize the importance of adaptive strategies, randomized allocation, and anticipatory defense in environments where attack and defense capabilities are uneven and constantly evolving. Min et al. apply a Colonel Blotto formulation to cloud storage security, showing that reinforcement learning–based CPU allocation improves both protection levels and overall system utility~\cite{min2018defense}. Ferdowsi et al. investigate cyber-physical systems in smart cities, demonstrating that when attackers lack knowledge of interdependencies, defenders can achieve a significant advantage~\cite{ferdowsi2017colonel}. Gupta et al. extend the model to a three-stage, three-player game in which coalitions may form, resources can be transferred, and battlefields modified, showing that coalition strategies can maximize defensive payoff under certain conditions\cite{gupta2014three}.

\subsection{Our approach}

To the best of our knowledge, we are the first to utilize a Colonel Blotto model for social engineering attacks and related preventive defensive efforts grounded in Routine Activity Theory.

Integrating RAT and the Blotto game framework requires aligning their underlying assumptions regarding actors, resources, and decision making. At a conceptual level, the two approaches are complementary. RAT explains why certain targets become attractive and vulnerable, while the Colonel Blotto model formalizes how strategic actors allocate scarce resources. The joint use of the two frameworks can be understood through their core elements. According to RAT crime occurs when a motivated offender encounters a suitable target in the absence of capable guardianship. These elements map naturally onto the Blotto model. The motivated offender corresponds to the attacking player. The suitable target is represented as a set of battlefields with heterogeneous values. Capable guardianship is reflected in the defensive allocation of resources. The VIVA dimensions further refine this mapping by providing a structured way to parameterize target characteristics seeding game instances with real-world data.

On the other hand, RAT is primarily descriptive and emphasizes situational opportunities, while assuming opportunistic offender behavior. In contrast, the Colonel Blotto model assumes fully strategic and rational actors who optimize their resource allocation under given informational conditions. This may distort reality because real world cybercriminals do not necessarily follow equilibrium strategies. These differences can be reconciled by interpreting the Blotto model as an approximation layer built on RAT's empirically grounded inputs such as target attractiveness, exposure, and the level of protection.

We deliberately focus on minimally sufficient models that i) preserve analytical transparency and computational tractability, ii) allow for parametrization based on real-world data, and iii) still produce actionable insights for policymakers and organizational decision-makers.

%% file: content/4_basegame.tex
\section{Game 1: Attackers vs. government}
\label{sec:basegame}

\subsection{Players}

Our first game models the interaction between attackers and governments. As argued in Section~\ref{sec:background}, a large fraction of social engineering attacks target the general public, and (aging) societies require particular attention and tailored prevention strategies. Since ensuring population cybersecurity preparedness falls primarily under the responsibility of the government, the defending player in our Colonel Blotto model is the nation-state. Cybercriminals may operate individually or as part of organized groups, and may reside either within the country or in various regions around the world. For the purposes of our analysis, we conceptualize them as a single large adversary posing a threat to the nation-state and its citizens. 

\subsection{Battlefields}

The different potential types of fraud constitute the battlefields. We narrow down the original 8 types defined in recent regional reports~\cite{enisa2024,ic3report2024} to 5 that have been sufficiently analyzed in the 2022 Cross-Sectional Survey of Cybercriminology~\cite{Parti2025}. This instantiation of cross-sectional online polls, started in 2019, focuses on cyber offending, cyber victimization, and criminology, while demographic and other computer-use questions are also included. The 2022 dataset includes three separate samples from different countries: the United States, Hungary, and Finland, tallying a total of $4,200$ participants. The survey sample represents the populations of all participating countries by age and sex, thereby providing an accurate reflection of the populations of the participating countries. 

The size of the battlefields is determined by the aggregate involvement of the citizens in the given type of fraud in the given country. We establish the level of involvement based on the survey questions that directly addressed this aspect:
\begin{enumerate}
\item \textbf{Tech support scam. }Tech support scams generally follow a highly consistent pattern, involving requests for remote access and subsequent demands for payment, making \textit{Q47} a pertinent measure of respondents’ exposure to this type of fraud.

\textit{Some scammers call people pretending they are from an  IT company or personnel, ask to allow remote access to  the computer, and once they are given access, they lock  the owner out. Then they ask for credit card details to  repair the owner’s computer. In the past 12 months, did you witness the above scenario?} 

\item \textbf{Grandparent/Relative Impersonation. } Grandparent/Relative Impersonation scams, in which fraudsters pose as relatives in need of urgent financial help, follow a recognizable pattern, making \textit{Q48} a useful indicator of the respondents’ exposure to this type of deception.

\textit{Some scammers call people pretending they are their grandchildren, asking for money to solve some
unexpected financial problem (overdue rent, payment for car repairs, bail, etc.). At the same time, the caller begs the grandparent ``please don’t tell my parents.'' In the past 12 months, have you got a call from such a scammer?}
 
\item \textbf{Government Impersonation. }Impersonation attacks, in which fraudsters pose as legitimate companies or government institutions to prompt recipients to update or verify personal information, exhibit a consistent pattern, making \textit{Q49}  a useful indicator of respondents’ exposure to this form of fraud.

\textit{In the past 12 months, have you received email from a legitimate company or institution (e.g. IRS, bank, etc.) asking you to ``update'' or ``verify'' your personal information via email or on a website provided by the email?}

\item \textbf{Investment scam. }Investment scams, in which individuals are asked to send money with the promise of receiving a larger return at the end of a cycle, follow a well-defined pattern. Therefore, \textit{Q50} serves as an effective indicator of respondents’ exposure to this type of fraud.

\textit{In the past 12 months, have you received a call or email according to which you would have been required to  send money to someone so that at the end of the cycle (when everyone pays a certain amount of dollar) you get back a greater amount of money?}

\item \textbf{Confidence/Romance Fraud. }Romance scams, in which individuals met through online dating platforms request money or other financial contributions under various pretexts, follow a consistent pattern, making \textit{Q51} an effective indicator of respondents’ exposure to this type of fraud.

\textit{In the past 12 months, have you been asked by someone you met on an online dating platform to send them money or other donations (e.g. plane tickets, travel expenses, etc.) or finances (e.g. pay for surgery or other medical expenses, pay custom fees to retrieve something, pay off gambling debts, pay for visa or other of financial travel documents, reload cards or gift cards)?}     
\end{enumerate}

As all listed questions are yes-or-no questions, simply counting the ratio of positive answers determines the relative size of each battlefield. Note that in order to work only with integer values i) the aggregated size of the 5 battlefields is normalized to $T = 20$, and ii) the individual sizes are first floored then incremented to the next integer in the order of decreasing fractional parts until reaching $T$. Furthermore, although the sizes of the battlefields differ, we assume that losing any of them results in the same loss to the defender, i.e., their value is considered equal. While this may be a simplification, a state should ideally defend each potential victim group similarly, justifying the assumption.\\

\noindent\textbf{Note on cybercrime surveys. }We are aware that survey-based cybercrime research is frequently criticized for potential inaccuracies stemming from self-report bias, non-probability sampling, rare-event measurement, and inflated prevalence estimates, particularly when surveys are used to generate national loss figures or population-wide rates~\cite{florencio2012sex, woods2022reviewing}. These critiques are well-founded when survey data are treated as substitutes for official statistics. However, as demonstrated by Parti et al.~\cite{parti2024perspectives}, such limitations do not invalidate the use of paid panel surveys for theory-driven analyses focused on behavioral correlates and relative risk rather than absolute prevalence. In the survey in question~\cite{Parti2025}, several design features were implemented to mitigate known sources of bias. U.S. data were collected through Dynata International, a professional panel vendor employing quota-based sampling to approximate the U.S. population by age, sex, and race, a strategy shown to produce samples closely aligned with U.S. Census benchmarks. The survey incorporated attention and speed checks to identify inattentive respondents, and researchers had no access to identifying information, ensuring respondent anonymity and reducing incentives for strategic misreporting. Importantly, Parti et al.~\cite{parti2024perspectives} provide empirical validation of Dynata-based samples by benchmarking them against U.S. Census data and demonstrating that Dynata surveys accurately predicted the 2020 U.S. presidential election—performing comparably to, and in some cases better than, established polling aggregators such as FiveThirtyEight. Together, this evidence indicates that while panel surveys have limitations for estimating absolute cybercrime rates, they provide a credible and methodologically appropriate foundation for testing criminological theories such as Routine Activity Theory, particularly when analytic goals center on patterns of exposure, online routines, and guardianship rather than population-level counts.

\subsection{Resource budget: number of troops}
\label{sec:game1_troops}

To properly reflect the characteristics of the defending country, the number of deployable defensive troops is derived from the nation-state’s cybersecurity expenditures, while the number of attacking troops is also primarily inferred from the country’s attributes. We intentionally introduced asymmetry into the model regarding defender and attacker capacities. Defender resources represent cybersecurity maturity levels, which typically evolve through discrete capability jumps due to budgetary, institutional, and operational constraints, making them suitable for categorical classification. In contrast, attacker capabilities are modeled incrementally, as cybercriminal groups operate with greater scalability, flexibility, and without regulatory constraints. Thus, this modeling feature can be interpreted as a dimension of the attacker's advantage. 

\noindent\textbf{Defensive troops. }For the defending nation-state, the number of troops is parameterized based on the total resources allocated to preventive activities, with law enforcement, defense, and corporate cybersecurity expenditures jointly considered as cyber defense–related costs. We classify nations into low, medium, and high categories based on their cybersecurity expenditures. For European countries, the reference is provided by the map developed by the European Cybersecurity Competence Center~\cite{COcyberMap2025}. In categorizing European countries, we used the indicator \textit{Share of Cybersecurity Projects Budget per Million People}, as this normalizes the budget by population size, enabling the comparison of countries of different sizes. For the United States, as no official expenditure data was available, we relied on the Digital Society Project~\cite{dsp_cyber_capacity}; interestingly, a quick cross-country analysis on the Government Cybersecurity Capacity indicator shows good alignment with the actual cybersecurity spending in~\cite{COcyberMap2025}. 

\begin{itemize}
\item \textbf{Low. }Countries were classified into the low category if their cybersecurity expenditure was below EUR 5 million per million inhabitants. Examples include Hungary, Poland, Bulgaria, and Serbia. Such countries are assigned 5 troops.

\item \textbf{Medium. }Countries were classified into the medium category if their cybersecurity expenditure was between EUR 5 million and EUR 10 million per million inhabitants. This category includes Germany, Sweden, Latvia, Lithuania, and others. Such countries receive 10 troops.

\item \textbf{High. }Countries belong to the high category if their cybersecurity expenditure exceeds €10 million per million inhabitants. Among the highest spenders are countries such as Finland,  Estonia, Norway, Cyprus, and Ireland, and the United States. Such countries receive 15 troops.
\end{itemize}

%%%%%%%attacking troops
\noindent\textbf{Attacking troops. }In Section~\ref{sec:background}, we introduced the concept of VIVA, which represents the variables Visibility, Inertia, Value, and Accessibility from the offenders’ perspective. We argue that the VIVA framework can be used to evaluate countries as aggregate target environments, thereby allowing us to determine the \emph{extent to which a given country attracts cyber offenders}; consequently, a country’s indicators determine the level of attacking capacity it has to face. 
%%%revision
We acknowledge that VIVA was originally developed as a micro-situational framework for assessing target suitability, and we do not treat an entire country as a single homogeneous victim. Rather, in this model, VIVA is used ecologically to characterize the aggregate attractiveness and reachability of a national target pool in a population-level prevention setting. This distinction is important because our unit of defense in Game 1 is the nation-state, not the individual citizen. Accordingly, country-level indicators such as social media penetration, cybersecurity maturity, income level, linguistic reach, and internet use are not intended to replace micro-level victimology; instead, they serve as proxies for how visible, accessible, valuable, and resistant a country’s population is to large-scale social engineering campaigns. This adaptation is consistent with prior cybercrime research showing that routine activity concepts can be examined at both individual and higher levels of aggregation, including country-level and multilevel analyses of victimization and guardianship~\cite{kigerl2012routine,leukfeldt2016applying,10.1093/bjc/azv011}. We therefore use VIVA here as a structured macro-approximation of offender target-pool selection, while recognizing that micro-situational heterogeneity remains crucial and is better captured in our organization-level model in Section~\ref{sec:extended}.
%%%end revision
We deliberately capped the number of attacking troops at $20$: together with country characteristics, this choice allows us to analyze both near-symmetric and asymmetric scenarios in terms od offense-defense balance.

Each VIVA variable is evaluated on a five-point scale (1–5) according to the following criteria.

\noindent\textbf{Visibility}. In order to assess how visible a country and its population are online, we based our evaluation on social media usage, drawing data from the World Population Review database~\cite{worldpopulationreview_smu2025}. Social media generates vast amounts of data and significantly increases individuals’ exposure to fraud and various types of attacks. From the database, we used the absolute number of social media users per country rather than percentages. Due to the characteristics of the online space, we aim to evaluate the likelihood that a user targeted by cybercriminals belongs to a given country. To determine the number of attackers attracted based on Visibility, we apply the conversion in Table \ref{tab:visibility_troops_b}.

\begin{table}[tb]
\centering
\begin{tabular}{|l|c|}
\hline
\textbf{Social Media Users} & \textbf{Attacking Troops} \\
\hline
less than 5M & 1 \\
\hline
between 5M and 20M & 2 \\
\hline
between 20M and 50M & 3 \\
\hline
between 50M and 100M & 4 \\
\hline
more than 100M & 5 \\
\hline
\end{tabular}
\caption{Attacking troops based on social media users}
\label{tab:visibility_troops_b}
\end{table}

\noindent\textbf{Inertia. }To determine Inertia, we utilize the ITU Global Cybersecurity Index (GCI), which is a composite score ranging from 0 to 100~\cite{itu_gci}, measuring the commitment of countries to cybersecurity at a global level. The GCI consists of five main pillars, each contributing to the overall score: Legal, Technical, Organizational, Capacity, and Cooperation Measures. We consider this comprehensive framework to provide sufficient depth for assessing a country’s level of vulnerability and determining the degree of Inertia it exhibits against cybercriminal activities. As a result, the number of attacking units is inversely proportional to the GCI score: the lower the GCI value, the more attackers are attracted to the country. The number of attacking units can be determined as a function of the GCI according to Table~\ref{tab:GCI}.

\begin{table}[tb]
\centering
\begin{tabular}{|l|c|}
\hline
\textbf{GCI -- Overall Score} & \textbf{Attacking Troops} \\
\hline
more than 95 & 1 \\
\hline
between 85 and 95 & 2 \\
\hline
between 55 and 85 & 3 \\
\hline
between 20 and 55 & 4 \\
\hline
less than 20 & 5 \\
\hline
\end{tabular}
\caption{Attacking troops based on the Global Cybersecurity Index}
\label{tab:GCI}
\end{table}

%%%revision new paragraph
In the context of social engineering, capable guardianship also includes the self-protective capacity of potential victims. We do not model individual cognition as a separate strategic player; however, self-protection remains conceptually central to our framework because deception-based attacks often succeed by exploiting bounded attention, time pressure, authority cues, and other cognitive vulnerabilities. From this perspective, awareness training, phishing education, callback verification, peer consultation, and organizational norms that encourage delay and double-checking can all be understood as mechanisms that strengthen guardianship by increasing users’ resistance to manipulation. This interpretation is consistent with prior cybercrime research that treats self-guardianship and protective online behavior as important dimensions of victimization risk, as well as with phishing research showing that self-regulation, self-efficacy, and behavioral training affect susceptibility to deception~\cite{leukfeldt2016applying,Vakhitova19052023,doi:10.1177/21582440231217720,RIBEIRO2024103558}. Thus, rather than excluding victim self-protection, our model incorporates it indirectly as a trainable and policy-relevant component of the broader guardianship environment.
%%%end revision

\noindent\textbf{Value. }From the perspective of cybercriminals, one of the most important variables is the value of the target. If the population of a target country generally possesses a higher level of economic well-being, the potential return from cybercrimes is correspondingly greater. To categorize countries, we utilize the World Bank classification, which divides countries into four categories based on GNI (Gross National Income) per capita~\cite{worldbank2025}. To align with other VIVA dimensions, we extended this classification with a fifth category, \textit{Upper-High Income}, with a threshold at least twice that of the High-Income category, resulting in a threshold of \$27,870 (see Table~\ref{tab:gni-countries}).

\begin{table}[tb]
\centering
\begin{tabular}{|c|c|c|}
\hline
\textbf{Income} & \textbf{Category} & \textbf{Attacking Troops} \\
\hline
\$1,135 or less & Low Income & 1 \\
\hline
between \$1,136 and \$4,495 & Lower-Middle Income & 2 \\
\hline
between \$4,496 and \$13,935 & Upper-Middle Income & 3 \\
\hline
between \$13,935 and \$27,870 & High Income & 4 \\
\hline
 \$27,870 or more & Upper-High Income & 5 \\
\hline

\end{tabular}
\caption{Number of attacking troops based on GNI}
\label{tab:gni-countries}
\end{table}

\noindent\textbf{Accessibility. }We have restricted the definition of accessibility to spoken language and Internet penetration among the population. Although the use of AI makes it easier to conduct high-quality native-language social engineering attacks, we contend that less widely spoken and more linguistically complex languages still increase the relative cost of attack, 
as AI-generated text and audio in such languages remain more easily detectable by native speakers. Countries whose official language(s) rank among the top ten most spoken languages~\cite{ethnologue200} are allocated two attacking troops, whereas countries outside this group receive only one additional attacking unit. We used the World Bank’s data to determine the percentage of the population using the Internet~\cite{worldbank_internet_usage}, and established cut‑offs at 50\%, 65\% and 80\%. The number of attacking troops assigned by the accessibility variable is determined according to Table~\ref{tab:accessibility_troops}.

\begin{table}[tb]
\centering
\begin{tabular}{|l|c|}
\hline
\textbf{Language} & \textbf{Attacking Troops } \\
\hline
Official language not among top 10 most spoken & 1 \\
\hline
Official language among top 10 most spoken & 2 \\
\hline
\textbf{Individuals using the Internet (\% of population)} & \textbf{Attacking Troops } \\
\hline
less than 50\% & 0 \\
\hline
between 50\% and 65\% & 1 \\
\hline
between 65\% and 80\% & 2 \\
\hline
more than 80\% & 3 \\
\hline
\end{tabular}
\caption{Attacking troops based on language and Internet usage}
\label{tab:accessibility_troops}
\end{table}

\subsection{Numerical evaluation}
%%%%%copied here
National policymakers must pre-select the areas to which they will allocate resources, as social engineering awareness programs typically last for several months or even years. Given this timeframe, attackers can vary their strategy; therefore, we are interested in a robust defensive allocation that yields the highest win probability across all potential offensive allocations.  

Our Python script runs a discrete, combinatorial simulation of the Colonel Blotto game\footnote{\url{https://anonymous.4open.science/r/Cybercrime-and-Prevention-Colonel-Blotto-in-Social-Engineering-073C}}. It aims to find the most effective ways for the defender to allocate resources. The script generates all possible soldier distributions for both the defender and the attacker given resource constraints on attacking and defending troops and the characteristics of the battlefields. Then, it compares these allocations to calculate how often each defensive strategy is successful against attacker strategies. Draw outcomes are seen as neutral events. Based on the results, the script ranks the defensive strategies by their win probabilities. Finally, it presents the ten most effective allocations in numerical form. The ten best defensive strategies are subsequently presented on a heatmap, where the X-axis represents the battlefields and the Y-axis indicates the index of each strategy along with its corresponding win probability. Each row corresponds to a distinct defensive strategy, while the color gradient, from light to dark, depicts the number of soldiers allocated to each battlefield. The script is subsequently run using data from Hungary, Finland, and the United States, and the resulting outcomes are analyzed. The characteristics of the three countries differ both in terms of budget and VIVA-related attributes, allowing for a comparative analysis of the different nations.

\subsubsection*{Scenario: Hungary}

\noindent\textbf{Battlefields. }As far as Hungarian survey respondents, 3.73\% reported having experienced tech support scams, 2.46\% encountered a grandparent/relative impersonation scam, 19.62\% experienced government impersonation, 9.99\% were approached with investment-related fraud, and 5.91\% reported confidence or romance scams. After normalization and integer adjustment, the resulting battlefield sizes are: $S = [2,\, 1,\, 9,\, 5,\, 3]$.

\noindent\textbf{Defending troops. }According to the COcyber Map~\cite{COcyberMap2025}, Hungary’s Share of Cybersecurity Projects Budget per Million People amounts to EUR 2.26 Million, placing the country into the bottom range among European countries. As a result, Hungary falls into the low category in terms of the number of defensive resources, possessing only \emph{5 defensive troops}.

\noindent\textbf{Attacking troops. }The country’s characteristics determine the number of attacking troops as follows: 7 million Hungarians use social media, which results in two attacking troops. Hungary’s Global Cybersecurity Index falls within the 85–95 range, adding two more. As a high-income country, Hungary attracts an additional four attacking soldiers. From a linguistic perspective, Hungarian is considered one of the most difficult languages, with an estimated 15 million speakers worldwide. In addition, the country is a leader in network infrastructure, with 94\% of the population using the Internet. These factors contribute an additional four enemies, bringing the total to \emph{12 attacking troops} overall.

\begin{figure}[tb]
	\centering
	\includegraphics[width=140mm, keepaspectratio]{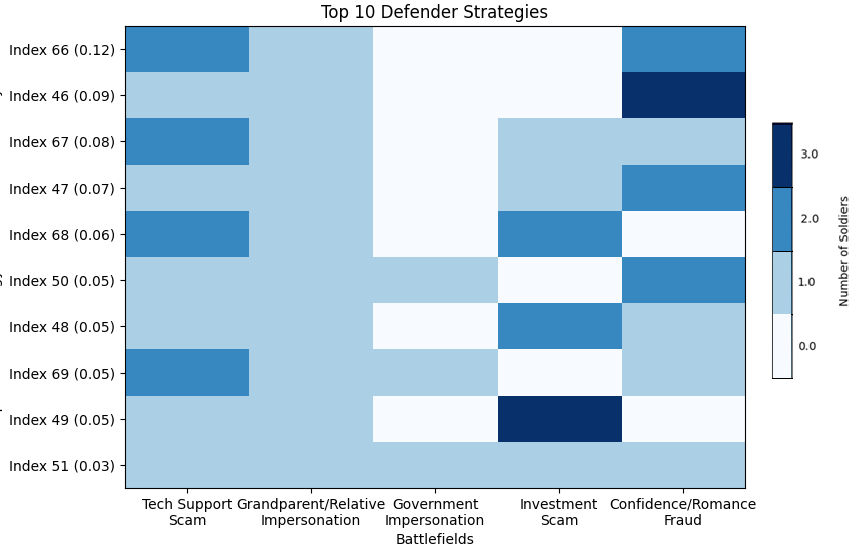}
	
	\caption{Heatmap of the 10 best defensive strategies of Hungary. The x-axis represents the battlefields, while the y-axis indicates the ranking of the best strategies and their corresponding win probability.}
	\label{fig:hunheat}
\end{figure}

\noindent\textbf{Results. }The simulations were run with the above parameters reflecting Hungary’s characteristics. We obtained the resulting Top 10 most effective defender strategies as depicted in Figure~\ref{fig:hunheat}.

Hungary faces an opponent with a significant advantage, as its characteristics attract a large number of attacking troops, while its relatively low expenditure results in limited defensive capacity. This imbalance is reflected in the low probability of a defensive victory, where the highest observed value reaches only 12\%. Government Impersonation represents the most prominent threat, whose magnitude alone can exceed the nation’s total defensive resources, making this battlefield one that Hungary is almost certain to lose. Consequently, the most effective strategy for the country is to focus on securing smaller battlefields, where achieving a win or draw is more feasible.

Trivially, the country has two potential pathways to improve its defensive position: increasing the number of defending troops or reducing the number of incoming attacking forces. Currently, Hungary’s Share of Cybersecurity Projects Budget per Million People amounts to EUR 2.26 million, placing it in the low spending category. To reach the medium category, the country would need to raise this figure to at least €5 million, corresponding to an estimated 10 defending units. This would mean increasing its total national cybersecurity budget by $\approx$ EUR 27 million. 

Hungary’s GCI sub-scores range between 85 and 95, with its weakest dimension being capacity development, which focuses on building cybersecurity expertise and awareness through education, training and research initiatives. If the country significantly increased its expenditures and directed most of the additional resources to capacity development, for example by training awareness specialists and providing awareness raising programs to individual users and vulnerable populations such as older individuals, it could double its defensive capabilities while facing a reduced number of attacking units.

\subsubsection*{Scenario: Finland}

\noindent\textbf{Battlefields. }For Finnish respondents, 12.15\% reported having experienced tech support scams, 3.31\% encountered a grandparent/relative impersonation scam, 26.38\% experienced government impersonation, 8.63\% faced investment-related fraud, and 6.76\% reported confidence or romance scams.
After normalization an integer adjustment, the resulting battlefield sizes are: $S = [4,\, 1,\, 9,\, 3,\, 3]$.

\noindent\textbf{Defending troops. }According to the COcyber Map~\cite{COcyberMap2025}, Finland’s Share of Cybersecurity Projects Budget per Million People amounts to EUR 19.02 million, placing the country in the seventh highest position among European nations. As a result, Finland falls into the high category in terms of defensive resources, possessing \emph{15 defensive troops}.

\noindent\textbf{Attacking troops. }The country’s characteristics determine the number of attacking troops as follows: 4.4 million Finns use social media, which results in one attacking troop. Finland’s Global Cybersecurity Index is more than 95, adding only one more. As an upper-high income country, Finland faces an additional five attacking soldiers. From a linguistic perspective, the Finnish language, similarly to Hungarian, is considered one of the most difficult and rarest languages in the world, with an estimated 6 million speakers worldwide. In addition, the country is a leader in digital society, with 94\% of the population using the Internet. These factors contribute an additional four enemies, bringing the total to \emph{11 attacking troops} overall.

\begin{figure}[tb]
	\centering
	\includegraphics[width=140mm, keepaspectratio]{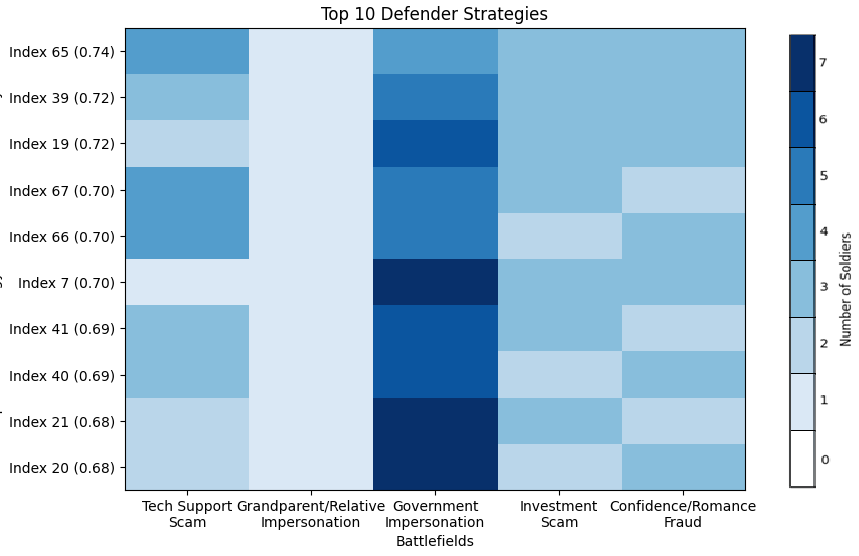}	
	\caption{Heatmap of the 10 best defensive strategies of Finland}
	\label{fig:finheat}
\end{figure}

\noindent\textbf{Results. }The simulations were run with the above parameters reflecting Finland’s characteristics. We obtained the resulting Top 10 most effective defender strategies as depicted in Figure~\ref{fig:finheat}.

The size of Finland’s battlefields exhibits notable similarities to those of Hungary. It is, however, evident that Finland’s substantial cybersecurity budget provides it with a dominant defensive advantage, reaching a win probability of $0.74$, allowing the country not only to allocate forces to smaller battlefields but also to achieve victories on the largest one. The ten strongest strategies show remarkable similarity, differing only by the movement of a single unit, which indicates a stable outcome. The consistent outcomes and high win probabilities demonstrate that Finland can operate with an optimal allocation of resources and is capable of maintaining a robust defense against cyberattacks. Overall, the country’s position corresponds to that of a stable, well-funded, and strategically mature defensive system.

\subsubsection*{Scenario: United States}

\noindent\textbf{Battlefields. }
For American respondents, 21.85\% reported having experienced tech support scams, 17.19\% encountered grandparent scams, 32.09\% experienced government impersonation, 18.12\% faced investment-related fraud, and 17.05\% reported confidence or romance scams. After normalization an integer adjustment, the resulting battlefield sizes are: $S = [4,\, 3,\, 6,\, 4,\, 3]$.

\noindent\textbf{Defending Troops. }Accurate data on the United States’ annual cybersecurity expenditures are not publicly available. However, according to the Digital Society Survey's data on the government's cybersecurity capacity~\cite{dsp_cyber_capacity}, the United States has consistently been a global leader between 2000 and 2025. Thus,
%2024 estimated range between \$80 and \$100 billion without the number for the Department of Defense. Given the nation’s population of 340.1 million, 
it can be concluded that the United States falls into the high-expenditure category, corresponding to an equivalent of \emph{15 defensive troops}.

\noindent\textbf{Attacking Troops. }The country’s characteristics determine the number of attacking troops as follows: 253 million US citizens use social media, which results in five attacking troops. America’s Global Cybersecurity Index is 100, adding only one more. As an upper-high income country, the US faces an additional five attacking soldiers. From a linguistic perspective, English is the most widely spoken language in the world, with approximately 1.5 billion speakers. Furthermore, the country is home to numerous global technology companies and large corporations, and 93\% of its population is an Internet user. These factors contribute an additional five enemies, bringing the total to \emph{16 attacking troops} overall.

\begin{figure}[tb]
	\centering
	\includegraphics[width=140mm, keepaspectratio]{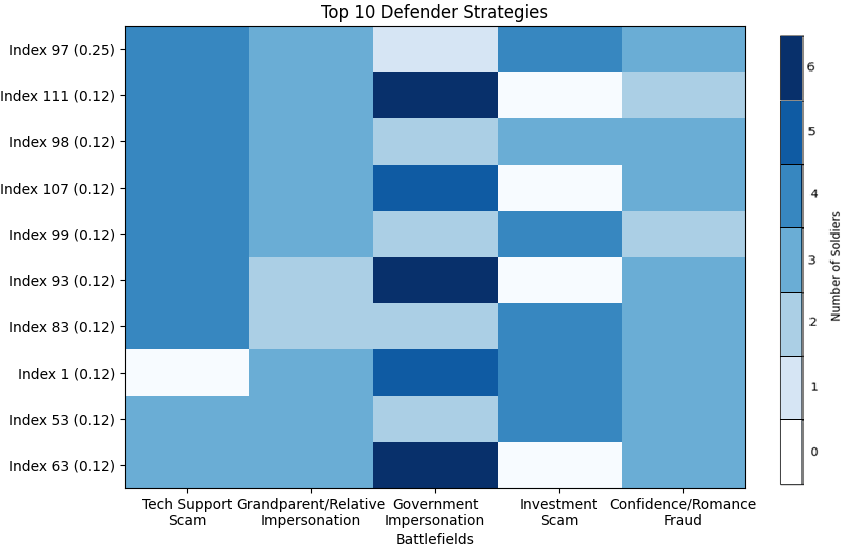}
	
	\caption{Heatmap of the 10 best defensive strategies of the United States}
	\label{fig:usaheat}
\end{figure}

\noindent\textbf{Results. }The simulations were run with the above parameters reflecting the United States’ characteristics. We obtained the resulting Top 10 most effective defender as depicted in Figure~\ref{fig:usaheat}.

Results clearly demonstrate that, despite its enormous budget, the United States can be considered a country with moderate defensive capabilities. Although virtually unlimited resources are available, the adversary’s offensive capacity remains extremely high. While in the cases of Hungary and Finland there is typically one distinctly dominant battlefield (government impersonation), in the United States the battlefields do not differ significantly in size. Most top strategies exhibit relatively low win probabilities of around 12\%, with considerable variation in the allocation of troops across battlefields. The most effective strategy achieves a 25\% win probability, though in this configuration the largest battlefield remains nearly undefended. The country managed to establish a balanced defense on the smaller battlefields relative to its capabilities, but due to the scale of the attacks, it cannot hold a dominant position.

\subsubsection*{Takeaway}

In designing the model, we sought to implement country-specific simulations that are as representative as possible given only publicly available information. Based on country-level statistics and a representative cybercrime survey having N$=1,400$ participants per country, we run numerical simulations for three different countries: Hungary, Finland, and the United States. In the case of Hungary, the results clearly indicate that a low cybersecurity budget constitutes a substantial disadvantage, and we identified the principal domain requiring future investment and capacity enhancement. Finland provides a representative positive example: substantial public funding for cybersecurity yields a clearly dominant defensive position relative to attackers. The situation in the United States is more nuanced: despite very large allocations to civilian (and, potentially, military) cyber defenses, the sheer volume of incoming attacks prevents the country from attaining a dominant position. Our model enables analysts to determine, for any country, the population-level position it occupies in defending against social engineering attacks, and to identify critical domains for future development.

%% file: content/5_extendedgame.tex
\section{Game 2: Attackers vs. organizations}
\label{sec:extended}
Building on Section~\ref{sec:basegame}, which analyzed population-level prevention from the perspective of nation-states, we now shift the focus to organizations.  The defender’s available resources are captured by the organization’s cybersecurity expenditure, whereas the attacker’s effective resources depend not only on their capabilities but also on attributes of the target organization that shape attack feasibility and expected returns.

%%%%
\subsection{Players}

In this model, the defending player is an organization, either a private firm or a public-sector entity, characterized by its size, budget constraints, and workforce composition. The attacking player represents either financially motivated cybercriminals or state-sponsored actors, whose objectives, capabilities, and operational methods may differ substantially.

\subsubsection*{Classifying organizations}

Organizations differ widely in their internal processes, workforce structure, and exposure to social-engineering threats. To support systematic comparison and parameterization, we group defenders into five high-level categories derived from Eurostat NACE Rev.~2~\cite{eurostat_nace_rev2_2008}. The classification emphasizes features that are particularly salient for social engineering, including training practices, role heterogeneity, physical and digital work environments, turnover, and the extent of external interaction.

\noindent\textbf{Critical infrastructure. }This category comprises organizations whose operations are essential to societal and economic functioning, including energy, water, telecommunications, transportation control, and financial infrastructure. Workforces are typically stable and comparatively well trained, with low turnover and long tenure. A central exposure arises from the breadth of deployed endpoints and customer-facing interfaces, which expands the attack surface and increases the potential value of unauthorized access. Insider risk is also non-trivial given privileged operational roles, while physical social engineering remains feasible due to large facilities and reliance on contractors and service providers.

\noindent\textbf{Industrial sector. }This category includes manufacturing, food processing, chemical production, automotive production, and large-scale logistics. Workforces are often dominated by operational and technical staff, whereas back-office functions such as finance and IT represent a smaller share. Business activity is primarily business-to-business, frequently embedded in geographically distributed supplier networks. This fragmentation increases process complexity and can weaken oversight, creating opportunities for adversaries to exploit procurement channels, impersonate partners, or target administrative functions that mediate payments and access.

\noindent\textbf{High customer-interaction services. }This category includes retail, customer service centers, bank branches, and entertainment services. It is commonly characterized by high turnover, part-time employment, and short onboarding cycles, often combined with sustained workload and frequent customer contact. These conditions increase susceptibility to manipulation, particularly for phishing, credential harvesting, and attempts at unauthorized physical access. Frontline employees typically have limited but non-negligible system access, and the frequent involvement of external vendors can introduce additional trust dependencies.

\noindent\textbf{Knowledge-based sectors. }This category includes IT firms, law offices, research institutes, and consultancies. Employees are highly skilled and operate in heavily digitized environments, frequently under hybrid or fully remote arrangements, sometimes with permissive device policies. In this setting, physical access is less central, but the value of digital assets (e.g., intellectual property, client data, privileged communications, and cloud resources) is high. Critical knowledge and access rights may be concentrated in small teams or individuals, making targeted social engineering attacks particularly attractive.

\noindent\textbf{Government services. }This category encompasses ministries, agencies, municipal administrations, law enforcement bodies, and public healthcare networks. These organizations routinely process large volumes of sensitive personal and operational information, often within hierarchical structures and bureaucratic workflows. Exposure arises both through digital systems and through the high frequency of in-person interactions with citizens, contractors, and external personnel. Resource constraints and procurement heterogeneity may lead to uneven technical baselines across offices, while time pressure and staff workload can amplify vulnerability to deception-based attacks.
\subsection{Battlefields}

To capture the organizational setting, we revise the battlefield structure to allow for heterogeneous targets and non-uniform payoffs. In practice, organizational attack surfaces differ in both size and value due to variation in internal processes, role structures, and operational priorities. Accordingly, battlefields in our model represent distinct social-engineering pathways, each associated with a different strategic importance and exposure level.

We define eight battlefields based on widely observed attacker methods. The first four reflect techniques commonly used by cybercriminals~\cite{verizon2024dbir}, while the remaining four correspond to approaches frequently associated with state-sponsored actors~\cite{mitreattack}. Importantly, these categories are not mutually exclusive: both cybercriminal and state-sponsored groups may employ tactics typically attributed to the other, depending on opportunity, constraints, and objectives.

\begin{enumerate}
\item \textbf{User Deception. }Organizations are exposed not only through employees but also through the users of their services and the customers of their products. Large-scale deception of end users can generate Human Denial of Service (HDoS) pressure, degrade service quality, and damage trust and reputation. This battlefield therefore captures adversarial attempts to manipulate external-facing interactions at scale. Its size is driven by the number of users and the breadth of available user actions and interaction channels.

\item \textbf{Internal Identity Exploitation. }This battlefield represents the misuse of internal identities and privileges, where adversaries leverage compromised accounts to expand access and influence within the organization. It captures escalation dynamics that rely on organizational permission structures rather than technical exploitation alone. Its size and value depend on the prevalence of elevated privileges and the degree of role separation and access segmentation.

\item \textbf{Physical Access. }Physical social engineering targets the organization’s premises and in-person workflows, including attempts to enter facilities, approach staff, or obtain sensitive information through direct interaction. Exposure is higher for organizations with multiple sites, walk-in services, or distributed points of sale. The size and value of this battlefield are determined by the number of locations and the range of on-site services and access opportunities.

\item \textbf{Urgency-Based Attacks. }Urgency-based manipulation exploits time pressure to induce rapid, insufficiently scrutinized decisions, such as overriding controls or authorizing exceptional actions. This approach is particularly effective where service continuity and time-sensitive transactions are critical. The size and value of this battlefield are shaped by the strictness of service-level expectations and the operational criticality of the targeted process.

\item \textbf{Targeted Relationship Building. }This battlefield captures long-horizon attacks based on trust formation, such as seemingly legitimate research collaboration, partnership proposals, or professional engagement. The objective is to gradually influence key individuals and extract valuable information or access through established rapport. Its size and value depend on the number of high-leverage personnel, the sensitivity of their privileges, and organizational factors that affect susceptibility to sustained manipulation.

\item \textbf{Supply Chain Attack. }Supply-chain social engineering targets trusted third parties to reach the organization indirectly. By compromising or impersonating established partners, adversaries exploit pre-existing trust relationships to bypass scrutiny. The size and value of this battlefield depend on the number of suppliers and partners, as well as the criticality of their operational roles and access pathways.

\item \textbf{Authority Impersonation. }Authority impersonation involves issuing plausible requests or directives while posing as regulators, law enforcement, tax authorities, or other oversight bodies. This tactic leverages compliance incentives and hierarchical norms to elicit sensitive information or induce harmful actions. The size and value of this battlefield are driven by the organization’s regulatory exposure and the extent to which hierarchical structures increase compliance with authoritative-seeming instructions.

\item \textbf{Critical Operator Manipulation. }This attack targets individuals whose actions can directly disrupt operations, such as employees with privileged control over core services or infrastructure. A single coerced action by such operators may halt services or trigger cascading failures. The size and value of this battlefield depend on the number of critical operators and the extent to which their actions require peer validation, separation of duties, or multi-party approval.
\end{enumerate}
%%

%%%%
To characterize the consequences of losing at a given battlefield, we assign each battlefield two parameters: a \emph{ratio}, capturing the exposure of the relevant personnel, and a \emph{value}, capturing the expected severity of harm if an attack succeeds in that domain. The ratio is measured on an ordinal scale from 0 to 3 and reflects how frequently employees associated with the battlefield are placed in situations where they can be targeted, manipulated, or pressured. For instance, physical intrusion is largely irrelevant for organizations that operate fully remotely and do not maintain customer-facing premises, yielding a ratio of 0. By contrast, organizations with multiple sites and extensive in-person service delivery (e.g., regional offices, retail locations, or public-facing administrative counters) expose a larger share of their workforce to direct interpersonal contact and receive the maximum ratio of 3. The \emph{value} of each battlefield is measured on a scale from 1 to 4 and reflects the magnitude of potential losses conditional on successful exploitation, including financial damage, operational disruption, legal exposure, and reputational harm. For example, in an industrial firm such as a lumberyard, most employees may have limited or no access to critical digital systems, implying that many individual compromises are likely to produce only localized harm. However, a smaller subset of personnel—such as IT administrators, finance staff, human resources, and operational supervisors—possess privileges and process access that can enable high-impact outcomes, ranging from fraudulent payments and payroll manipulation to prolonged downtime or even existential business risk. In this sense, the value parameter captures the damage potential associated with compromising the organizational function represented by the battlefield, rather than the average access level of a typical employee. Finally, to keep the model tractable and consistent with common budgeting practices, we assume that allocating defensive resources (e.g., training and awareness interventions) incurs comparable marginal cost per employee, even though the expected loss associated with compromise can differ substantially across roles.

Table \ref{tab:value_ratio} presents the method for calculating the values and ratios of battlefields based on personnel involvement and potential harm.
\begin{table}[tb]
\centering
\begin{tabular}{|l|c|}
\hline
\textbf{Personnel involvement} & \textbf{Battlefield ratio} \\
\hline
No personnel affected & 0 \\
\hline
Less than 25\%  & 1 \\
\hline
Between 25\% and 50\% & 2 \\
\hline
More than 50\%  & 3 \\
\hline
\textbf{Potential damage} & \textbf{Battlefield value} \\
\hline
Low: Easily remediable inconveniences & 1 \\
\hline
Medium: Disruption of certain services & 2 \\
\hline
High: The organization sustains major damages & 3 \\
\hline
Critical: The organization becomes inoperable & 4 \\
\hline
\end{tabular}
\caption{Battlefield Ratio and Value}
\label{tab:value_ratio}
\end{table}

After determining the ratios (and values) of the battlefields, the size of the existing battlefields can be established similarly to Game 1. Note that in order to work with integer values i) the aggregate battlefield size is normalized to  $T = 24$, and ii) the individual sizes are first floored then incremented to the next integer in the order of decreasing fractional parts until reaching $T$. 

\subsection{Resource budget: number of troops}

In practice, there is no universal rule that prescribes how much an organization should invest in cybersecurity, and spending levels remain highly heterogeneous across sectors, jurisdictions, and organizational maturity. A Deloitte study reports that financial institutions in the United States devote, on average, approximately 10\% of their IT budgets to cybersecurity, with this share increasing to 15--20\% for critical infrastructure operators~\cite{bernard2020reshaping}. However, reliable estimates of the fraction of cybersecurity expenditure specifically dedicated to mitigating social engineering risk (e.g., training, awareness, and human-centric controls) remain scarce. We therefore assume that general cybersecurity spending implicitly incorporates investments in workforce preparedness and set the \emph{defender’s maximum budget to $20$ troops}.

To reflect the structural advantage of attackers, who can select targets, reuse infrastructure, and iterate rapidly, we allow the \emph{attacking player up to $24$ troops}. Consistent with Section~\ref{sec:game1_troops}, the attacker’s troop budget is operationalized using the VIVA framework, which interprets attack attractiveness as a function of the target’s \textit{Visibility}, \textit{Inertia}, \textit{Value}, and \textit{Accessibility}. Conceptually, VIVA serves as a scoring mechanism that maps observable organizational characteristics to expected attacker attention and effort. Each dimension contributes between 0 and 6 troops, yielding a maximum of $24$.

\noindent\textbf{Visibility. }Visibility captures how easily an organization can be identified, profiled, and monitored by adversaries. It reflects the organization’s digital footprint (e.g., web presence, social media activity, advertising, and recruitment) as well as its physical footprint (e.g., the number and geographic dispersion of sites). Higher visibility increases the likelihood of being selected for opportunistic campaigns and also supports more targeted social engineering by enabling adversaries to gather contextual information and identify plausible pretexts (Table~\ref{tab:visibility_troops}).

\begin{table}[tb]
\centering
\begin{tabular}{|l|c|}
\hline
\textbf{Online activity} & \textbf{Attacking troops} \\
\hline
No online presence & 0 \\
\hline
Has a website/social media but inactive & 1 \\
\hline
Has a website/social media and it is active & 2 \\
\hline
Online advertising and recruitment & 3 \\
\hline
\textbf{Physical visibility} & \textbf{Attacking troops} \\
\hline
No physical location & 0 \\
\hline
Less than 4 locations & 1 \\
\hline
More than 3 locations nationwide & 2 \\
\hline
Has international locations & 3 \\
\hline
\end{tabular}
\caption{Attacking troops attracted by Visibility}
\label{tab:visibility_troops}
\end{table}

\noindent\textbf{Inertia. }Inertia represents the degree of resistance an organization can impose against social engineering attempts, reflecting both formal constraints (regulatory obligations, sectoral standards) and informal organizational capacity (governance maturity, leadership attention, and workforce awareness). Higher inertia implies that attacks require greater effort to succeed, as employees are more likely to detect manipulation attempts and internal processes are more likely to enforce verification and escalation. Conversely, low inertia corresponds to weak procedural friction and limited training, which lowers the attacker’s cost of exploitation (Table~\ref{tab:inertia_troops_ex}).

\begin{table}[tb]
\centering
\begin{tabular}{|l|c|}
\hline
\textbf{Regulations and legal requirements} & \textbf{Attacking troops} \\
\hline
International regulatory frameworks and standards & 1 \\
\hline
National regulations and sector-specific standards & 2 \\
\hline
Local or national regulations & 3 \\
\hline
\textbf{Management’s security awareness} & \textbf{Attacking troops} \\
\hline
Mature cybersecurity program, CISO-level oversight & 1 \\
\hline
Policies in place, regular employee training & 2 \\
\hline
Basic policies exist, no employee training & 3 \\
\hline
\end{tabular}
\caption{Attacking troops drawn by Inertia}
\label{tab:inertia_troops_ex}
\end{table}

\noindent\textbf{Value. }Value captures the potential payoff of a successful attack. For market actors, this may be proxied by revenue and the scale of financial flows, while for public-sector and nonprofit entities it is more closely related to operational budgets, service criticality, and the societal impact of disruption. In addition to economic scale, value also depends on the sensitivity of data processed by the organization, since personal, financial, or classified information can be monetized directly, leveraged for extortion, or exploited for intelligence purposes. We operationalize this dimension using the European Commission’s enterprise-size recommendations~\cite{EC-2003-361} together with a data-sensitivity component (Table~\ref{tab:value_troops}).

\begin{table}[tb]
\centering
\begin{tabular}{|l|c|}
\hline
\textbf{Annual turnover} & \textbf{Attacking troops} \\
\hline
Small enterprise: less than €10 million & 1 \\
\hline
Medium enterprise: between €10 million and €50 million & 2 \\
\hline
Large enterprise: greater than €50 million & 3 \\
\hline
\textbf{Data Sensitivity} & \textbf{Attacking troops} \\
\hline
Internal operational data & 1 \\
\hline
Confidential personal or financial data & 2 \\
\hline
Highly sensitive or classified data & 3 \\
\hline
\end{tabular}
\caption{Attacking troops drawn by Value}
\label{tab:value_troops}
\end{table}

\noindent\textbf{Accessibility. }Accessibility reflects the extent to which adversaries can establish contact with employees and users, deliver persuasive stimuli, and maintain interaction long enough to complete an exploit. This includes the availability of online entry points (e.g., customer portals, interactive platforms, real-time communication tools) as well as offline opportunities (e.g., public-facing facilities, walk-in services, and emergency contact channels). High accessibility increases the feasible volume of attack attempts and broadens the range of social engineering techniques that can be deployed, from phishing and impersonation to in-person manipulation (Table~\ref{tab:accessibility_troops_ex}).

\begin{table}[tb]
\centering
\begin{tabular}{|l|c|}
\hline
\textbf{Online Services} & \textbf{Attacking troops} \\
\hline
No online presence & 0 \\
\hline
Limited functions, contact forms & 1 \\
\hline
Interactive platforms and customer portals & 2 \\
\hline
Real-time online communication tools & 3 \\
\hline
\textbf{Offline Services} & \textbf{Attacking troops} \\
\hline
No physical locations & 0 \\
\hline
Restricted opportunities for in-person contact & 1 \\
\hline
Multiple locations accessible to users or employees & 2 \\
\hline
Multiple nonstop locations for emergency contact & 3 \\
\hline
\end{tabular}
\caption{Attacking troops drawn by Accessibility}
\label{tab:accessibility_troops_ex}
\end{table}
\subsection{Numerical evaluation}

After parameterizing the model with organization-specific variables, we solve the resulting game computationally. The equilibrium allocation provides a tractable proxy for an organization’s relative preparedness against social engineering threats and enables counterfactual analysis of how marginal changes in security investment may shift exposure across attack vectors.

For the numerical evaluation, we selected five organizations from distinct sectors and derived all their parameters exclusively from open-source intelligence (OSINT). To mitigate identification risk, organization names and contextual details were anonymized while preserving the relevant structural characteristics. Simulations were implemented in Python\footnote{\url{https://anonymous.4open.science/r/Cybercrime-and-Prevention-Colonel-Blotto-in-Social-Engineering-073C}}. In each run, we sampled a large set of feasible pure strategies for both players and computed the corresponding payoff matrix, which specifies the outcome for each attacker-defender allocation pair. We then approximated a mixed-strategy Nash equilibrium using regret matching~\cite{hart2013simple}. Intuitively, regret matching iteratively updates the probability of selecting each strategy based on its \emph{positive regret}, i.e., how much better the player would have performed on average had that strategy been played more often in the past. Over repeated iterations, this adaptive process concentrates probability mass on strategies that consistently outperform alternatives, and the empirical distribution of play converges to an equilibrium in average behavior. The resulting mixed strategies therefore provide an interpretable estimate of how attackers and defenders are expected to allocate resources across battlefields under strategic interaction.

\subsubsection*{Scenario: BrightLine Energy Plc.}

BrightLine Energy Plc. (BE) is a regional electricity distribution company with exclusive rights to supply residential and commercial customers. The company maintains a basic operational website, while its social media presence has been inactive for several years. BE operates weekday customer service offices across the region (in settlements above 5{,}000 inhabitants), maintains multiple technical facilities, and provides 24/7 online and telephone emergency reporting. As a critical infrastructure operator, BE is subject to NIS2 security requirements, including annual mandatory employee training delivered by an external provider. The region hosts several large industrial sites, placing BE’s annual turnover above EUR 50M. BE processes sensitive consumption and billing data and retains these records under statutory obligations. The company coordinates its security posture with peer electricity distributors nationwide and allocates approximately 15\% of its IT budget to cybersecurity. These characteristics correspond to 15 defensive troops and 16 attacking troops.

\noindent\textbf{Battlefields. }Approximately 40\% of BE’s workforce is office-based: 20\% in customer-facing roles and 20\% in IT, finance, and legal functions. The remaining 60\% support the deployment, operation, and maintenance of the power network, with roughly 5\% assigned to operations control tasks. User deception may overload complaint and fault-reporting channels, reducing service quality and damaging reputation. Internal identity exploitation is primarily concentrated among office personnel and may enable payment fraud, data exfiltration, or disruptive lateral movement. Physical access remains relevant due to distributed offices and technical sites. Urgency-based attacks are salient in critical infrastructure settings, where time pressure around outages can induce procedural bypasses. Supply-chain risk is shaped by reliance on a national electricity distributor, and authority impersonation is credible due to oversight by the national energy authority. Table~\ref{tab:BE_size_ratio} summarizes the battlefield ratios and values; the resulting integer-adjusted battlefield sizes are:
$[2,\, 3,\, 2,\, 5,\, 3,\, 5,\, 3,\, 1]$.

\begin{table}[tb]
\centering
\begin{tabular}{|c|c|c|c|c|c|c|c|c|}
\hline
\textbf{Battlefield:} & \textbf{1} & \textbf{2} & \textbf{3} & \textbf{4} & \textbf{5} & \textbf{6} & \textbf{7} & \textbf{8} \\
\hline
\textbf{Ratio}: & 1 & 2 & 1 & 3 & 2 & 3 & 2 & 1 \\
\hline
\textbf{Value}: & 2 & 3 & 4 & 3 & 3 & 4 & 4 & 4 \\
\hline
\end{tabular}
\caption{BrightLine Energy Plc.'s battlefields}
\label{tab:BE_size_ratio}
\end{table}

\begin{figure}[tb]
	\centering
	\includegraphics[width=140mm, keepaspectratio]{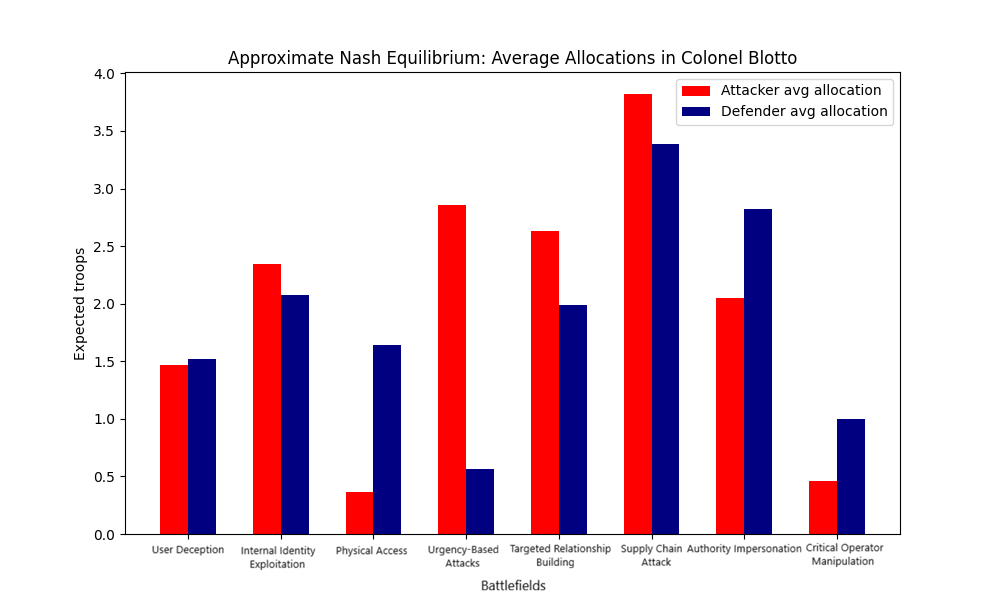}
	\caption{BrightLine Energy Plc.: approximate Nash equilibrium}
	\label{fig:BE_nash}
\end{figure}

\noindent\textbf{Results. }BE deploys \emph{15 defensive troops against 16 attacking troops}. Figure~\ref{fig:BE_nash} reports the approximate Nash equilibrium. The defender concentrates relatively more on battlefields \#1, \#3, \#7, and \#8, while the attacker emphasizes the remaining four. Neither player leaves any battlefield unallocated. The total value of defender-preferred battlefields is 13, compared to 14 for the attacker-preferred set, indicating that the attacker can impose greater expected harm while contesting a comparable number of battlefields. Within the model, BE can improve its position by modestly increasing defensive resources or by reducing attacker advantage through stronger security governance (e.g., improved management preparedness and CISO-level oversight).

\subsubsection*{Scenario: Oak \& Pine Co.}

Oak \& Pine Co. (OP) is a small, family-owned wood processing company with 30 years of operation. The firm has no website or social media presence. One site handles processing, while the other is dedicated to storage and corporate client sales. OP is subject only to local municipal regulations and a limited set of national laws, and management has not implemented formal employee training. Annual revenue does not exceed EUR 10M, and only essential operational data are processed. Sales occur exclusively with three long-term corporate partners, and the company provides no online services. At the request of its largest client, OP dedicates 5\% of its IT budget to cybersecurity initiatives. These characteristics correspond to \emph{5 defensive and 10 attacking troops}.

\noindent\textbf{Battlefields. }Most employees (85\%) are manual workers or maintenance staff, while the remaining 15\% manage administrative, IT, financial, and legal functions. Office personnel perform diverse duties, so internal identity exploitation or physical intrusion could produce significant disruption within administrative systems. Urgency-based attacks are largely irrelevant due to the predictable nature of operations. Targeted deception would primarily affect office staff; while a successful attack could incur notable losses, core wood-processing operations remain largely uninterrupted. Supply-chain compromise could interrupt the delivery of essential materials. Impersonation of authorities may lead only to minor regulatory or operational disruption. Finally, deception of the plant manager—the critical operator—could halt the entire facility. Table~\ref{tab:OP_size_ratio} presents battlefield ratios and values; the integer-adjusted battlefield sizes are $[2,\, 2,\, 6,\, 2,\, 2,\, 6,\, 2,\, 2]$.

\begin{table}[tb]
\centering
\begin{tabular}{|c|c|c|c|c|c|c|c|c|}
\hline
\textbf{Battlefield:} & \textbf{1} & \textbf{2} & \textbf{3} & \textbf{4} & \textbf{5} & \textbf{6} & \textbf{7} & \textbf{8} \\
\hline
\textbf{Ratio}: & 1 & 1 & 3 & 1 & 1 & 3 & 1 & 1 \\
\hline
\textbf{Value}: & 1 & 4 & 3 & 1 & 3 & 4 & 2 & 4 \\
\hline
\end{tabular}
\caption{Oak \& Pine Co.'s battlefields}
\label{tab:OP_size_ratio}
\end{table}

\begin{figure}[!tb]
	\centering
	\includegraphics[width=140mm, keepaspectratio]{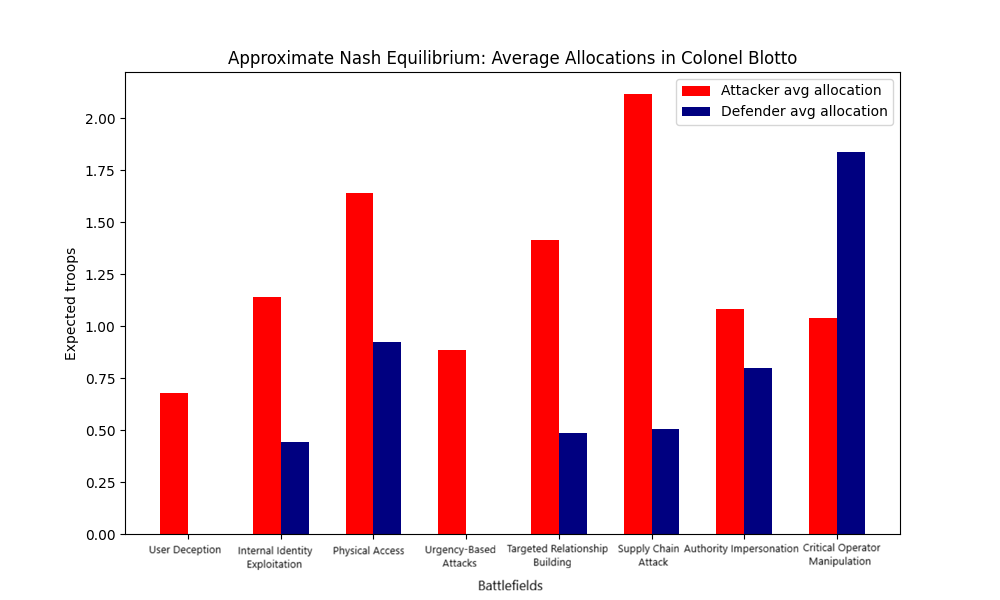}
	\caption{Oak \& Pine Co.: approximate Nash equilibrium}
	\label{fig:OP_nash}
\end{figure}

\noindent\textbf{Results. }With only 5 defensive units, OP is under-resourced relative to 10 attacking troops. As shown in Figure~\ref{fig:OP_nash}, attackers dominate nearly all battlefields on average, with two low-value battlefields left completely unprotected. The defender gains a marginal advantage only in protecting the critical operator, while internal identity exploitation remains vulnerable. Improving the organization’s cybersecurity posture would require at least doubling the defensive resources and strategically allocating them, primarily via formalized security policies and structured employee training. Doing so would also reduce the number of attacking units attracted by the organization’s characteristics.

\subsubsection*{Scenario: GlobalShield Insurance Inc.}

GlobalShield Insurance Inc. (GI) is an international insurance company operating in multiple countries. The firm actively advertises, recruits, and shares results on social media. Within its home country, GI maintains 150 offices and 30 international locations, serving clients on weekdays and providing 24/7 online and telephone support for claims and emergencies. Employee training is mandatory, cybersecurity is overseen by a CISO, and the company complies with all relevant international standards. Annual turnover exceeds EUR 50M, and 20\% of its IT budget is devoted to cybersecurity. These characteristics correspond to \emph{20 defensive and 19 attacking troops}.

\noindent\textbf{Battlefields. }Approximately 55\% of employees interact directly with clients, while 45\% perform back-office tasks. A large number of deceived users could overload staff, potentially disrupting service. Nearly all employees handle sensitive personal data, making internal identity exploitation and unauthorized physical access highly consequential. Urgency-based attacks primarily affect back-office staff but are mitigated by strict procedures. Customer-facing employees are highly exposed due to public contact information and social media activity. Supply chain compromise could disrupt critical processes, while authority impersonation may affect only partial operations. Manipulation of critical operators, especially IT personnel, could incapacitate the organization. Table~\ref{tab:GS_size_ratio} shows battlefield ratios and values; integer-adjusted battlefield sizes are $[4,\, 4,\, 3,\, 3,\, 4,\, 3,\, 1,\, 2]$.

\begin{table}[tb]
\centering
\begin{tabular}{|c|c|c|c|c|c|c|c|c|}
\hline
\textbf{Battlefield:} & \textbf{1} & \textbf{2} & \textbf{3} & \textbf{4} & \textbf{5} & \textbf{6} & \textbf{7} & \textbf{8} \\
\hline
\textbf{Ratio}: & 3 & 3 & 2 & 2 & 3 & 2 & 1 & 1 \\
\hline
\textbf{Value}: & 3 & 3 & 3 & 2 & 3 & 3 & 2 & 4 \\
\hline
\end{tabular}
\caption{GlobalShield Insurance Inc.'s battlefields}
\label{tab:GS_size_ratio}
\end{table}

\begin{figure}[tb]
	\centering
	\includegraphics[width=140mm, keepaspectratio]{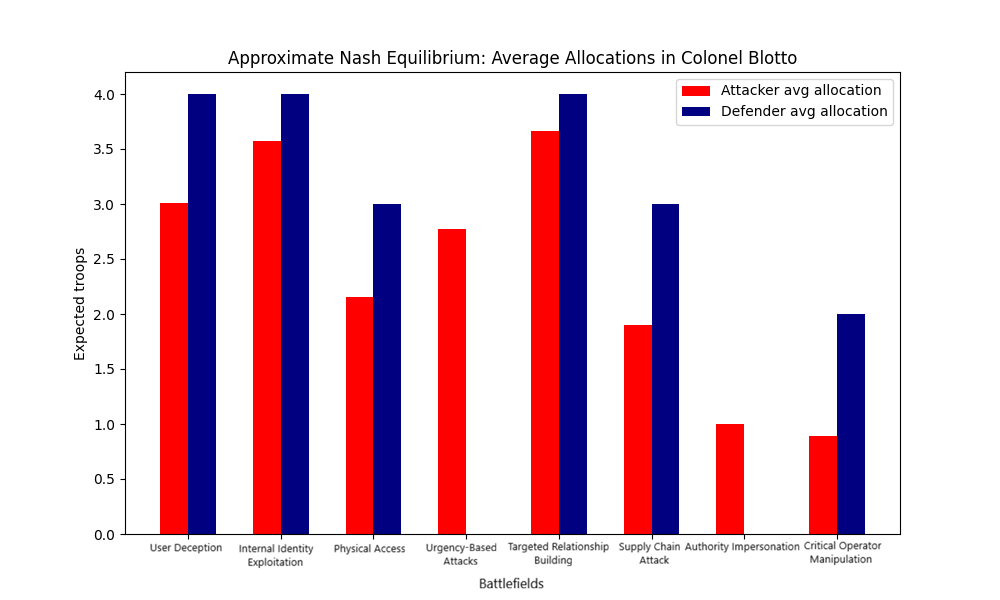}
	\caption{GlobalShield Insurance Inc.: approximate Nash equilibrium}
	\label{fig:GS_nash}
\end{figure}

\noindent\textbf{Results. }With 20 defensive units against 19 attacking units, GI achieves superiority on six battlefields on average, leaving two less critical battlefields undefended (Figure~\ref{fig:GS_nash}). The defense effectively protects the most valuable targets. Additional cybersecurity spending would yield limited gains; reducing attacker units would require limiting social media presence, altering recruitment, or modifying customer interaction channels, which could negatively impact revenue and market share. Maintaining the current cybersecurity investment is therefore the optimal strategy.

\subsubsection*{Scenario: ForwardThink Ltd.}

ForwardThink Ltd. (FT) is a small think tank supporting strategic decisions for corporations and government agencies. The firm maintains a minimal website with a central contact email, has no social media presence, and does not engage in advertising. Employees work primarily remotely, meeting clients or in co-working spaces only as needed. The organization handles highly sensitive client data and complies with strict national technology directives, though no additional security training is provided. Due to the small number of annual projects (maximum four) and employees, annual turnover does not exceed EUR 10M. Cybersecurity expenditures amount to 10\% of the IT budget, corresponding to \emph{10 defensive and 12 attacking troops}.

\noindent\textbf{Battlefields. }FT has no traditional users and operates without dedicated offices. Only a few employees manage finance and legal operations, and the IT infrastructure is outsourced, with just two internal system administrators whose compromise could incapacitate the organization. Internal identity exploitation could expose sensitive project information. Given the small project volume and prearranged schedules, urgency-based attacks are unlikely to cause major disruptions. Almost all personnel handle classified information, making them potential targets for highly consequential attacks, while authority impersonation against the legal officer would likely have limited operational impact. Table~\ref{tab:FT_size_ratio} summarizes battlefield ratios and values; integer-adjusted battlefield sizes are $[5,\, 2,\, 5,\, 2,\, 2,\, 2]$.

\begin{table}[tb]
\centering
\begin{tabular}{|c|c|c|c|c|c|c|c|c|}
\hline
\textbf{Battlefield:} & \textbf{1} & \textbf{2} & \textbf{3} & \textbf{4} & \textbf{5} & \textbf{6} & \textbf{7} & \textbf{8} \\
\hline
\textbf{Ratio}: & 0 & 3 & 0 & 1 & 3 & 1 & 1 & 1 \\
\hline
\textbf{Value}: & 0 & 3 & 0 & 1 & 4 & 4 & 2 & 4 \\
\hline
\end{tabular}
\caption{ForwardThink Ltd.'s battlefields}
\label{tab:FT_size_ratio}
\end{table}

\begin{figure}[tb]
	\centering
	\includegraphics[width=140mm, keepaspectratio]{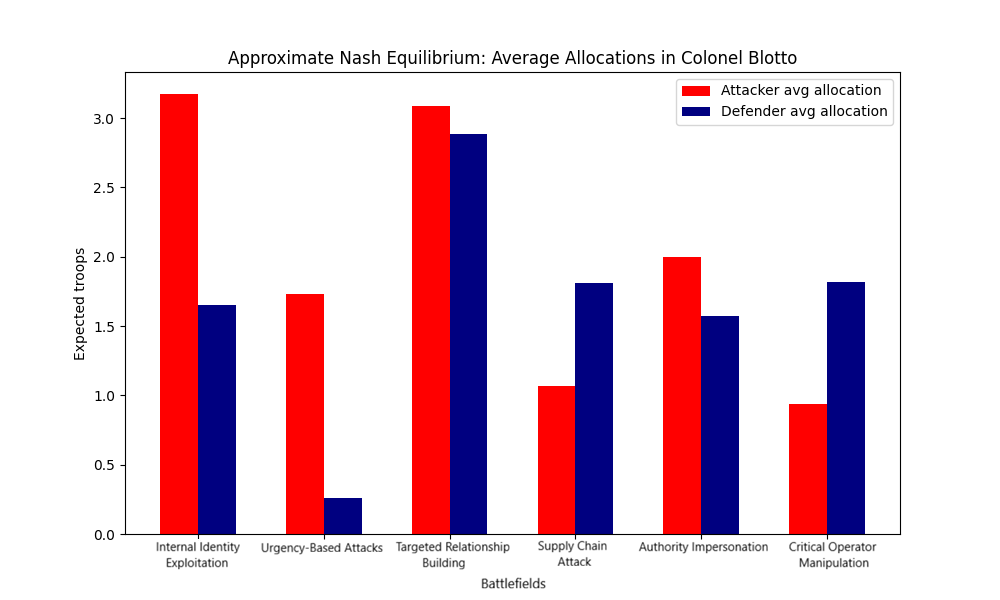}
	\caption{ForwardThink Ltd.: approximate Nash equilibrium}
	\label{fig:FT_nash}
\end{figure}

\noindent\textbf{Results. }FT defends six relevant battlefields with 10 defensive troops against 12 attacking troops. Figure~\ref{fig:FT_nash} shows the defense is insufficient, achieving superiority on only two smaller critical battlefields (combined value 8), while the attacker dominates the remaining four (total value 10). A modest increase in cybersecurity spending and targeted employee training would both increase the number of defensive units and reduce the number of attacking units, potentially equalizing forces. Given the handling of highly sensitive data, appointing a CISO would further strengthen the company’s defensive posture, enabling it to operate from a position of advantage within the model.

\subsubsection*{Scenario: CivicWatch}

CivicWatch (CW) is a municipal organization managing urban public spaces and supporting state police operations. High staff turnover necessitates continuous recruitment via social media. The organization operates three offices in outer districts, providing 24/7 complaint and crime reporting services and a telephone dispatch line. Following previous cyberattacks, cybersecurity has been prioritized, with compliance to all national regulations. Annual turnover does not exceed EUR 10M, and employees are authorized to issue fines and perform enforcement tasks, generating highly sensitive data. An online citizen portal has been established, and 12\% of the limited IT budget is allocated to cybersecurity. The organization corresponds to \emph{12 defensive and 16 attacking troops}.

\noindent\textbf{Battlefields. }City residents represent CW’s user base, and mass deception could overload staff, disrupting services. Most employees have access to personal data systems, making internal identity exploitation a significant legal and reputational risk. Physical access to offices would jeopardize surveillance and communication systems. Urgency-based attacks could strain available personnel, potentially causing outages. Many employees engage directly with residents, and access to personal data is common. The primary supplier is the municipality, and manipulation of supply chains or critical operators could lead to operational failures. CivicWatch operates under municipal and police authority, so false instructions from these entities could cause severe disruption. Table~\ref{tab:CW_size_ratio} presents battlefield ratios and values; integer-adjusted battlefield sizes are $[4,\, 4,\, 4,\, 4,\, 4 ,\, 1 ,\, 2 ,\, 1]$.

\begin{table}[tb]
\centering
\begin{tabular}{|c|c|c|c|c|c|c|c|c|}
\hline
\textbf{Battlefield:} & \textbf{1} & \textbf{2} & \textbf{3} & \textbf{4} & \textbf{5} & \textbf{6} & \textbf{7} & \textbf{8} \\
\hline
\textbf{Ratio}: & 3 & 3 & 3 & 3 & 3 & 1 & 2 & 1 \\
\hline
\textbf{Value}: & 2 & 3 & 4 & 2 & 3 & 4 & 4 & 4 \\
\hline
\end{tabular}
\caption{CivicWatch's battlefields}
\label{tab:CW_size_ratio}
\end{table}

\begin{figure}[tb]
	\centering
	\includegraphics[width=140mm, keepaspectratio]{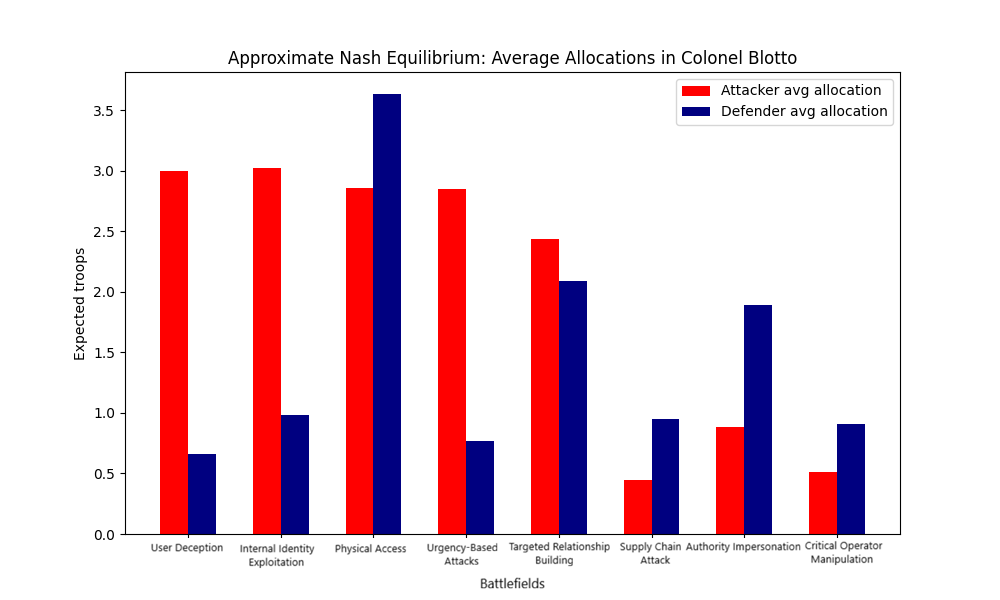}
	\caption{CivicWatch: approximate Nash equilibrium}
	\label{fig:CW_nash}
\end{figure}

\noindent\textbf{Results. }CW deploys 12 defensive troops against 16 attackers, slightly below average. Figure~\ref{fig:CW_nash} shows that defenders hold an advantage on only four smaller-value battlefields, while attackers dominate the remaining four critical ones. Given the organization’s responsibilities and structure, the number of attacking units cannot be reduced and must be treated as constant. Half of CW’s battlefields are critical, so insufficient protection could lead to operational shutdowns. Strengthening defense requires a substantial increase in cybersecurity spending, potentially 40-50\%, to allocate at least 17 defensive units. Such an investment would secure nearly all critical battlefields and provide the defender with a clear advantage over the attacker.

\subsubsection*{Takeaway}

The scenario analyses of five fictional organizations illustrate how size, budget, and workforce composition jointly shape defense against social-engineering attacks. BrightLine Energy Plc., with strong investment and regulatory compliance, achieves near parity with attackers, though additional spending or management improvements could enhance resilience. Oak \& Pine Co., a small family business with minimal cybersecurity, is highly vulnerable, highlighting the importance of even modest investment and training. GlobalShield Insurance Inc., a well-resourced international insurer, maintains a dominant defensive posture, demonstrating that structured programs can counter diverse attacks. ForwardThink Ltd., a small consultancy, has limited defenses but could benefit from targeted increases in investment and training. CivicWatch, a municipal organization with moderate resources, is vulnerable on half of its battlefields, requiring substantial cybersecurity spending to secure an advantage.

These cases reveal three key insights: (i) the ratio of defensive to attacking units is critical; (ii) organizational structure, employee roles, and operational dependencies shape battlefield values and exposure; and (iii) modest, strategically directed investments can significantly improve defense. The model provides a tractable framework to assess organizational preparedness and prioritize interventions to reduce risk.

%% file: content/6_conclusion.tex
\section{Conclusion}
\label{sec:conclusion}
While technological defenses continue to advance, the human factor remains a persistent vulnerability that can undermine even the most sophisticated security infrastructure. Our work contributes to the intersection of cybercrime prevention, behavioral security, and strategic decision-making by providing theoretically grounded, data-driven frameworks that enable both policymakers and organizational leaders to allocate defensive resources more effectively.

We presented two Colonel Blotto game models grounded in Routine Activity Theory and operationalized through the VIVA framework (Value, Inertia, Visibility, Accessibility). The first model examines population-level defense from the perspective of nation-states, demonstrating how countries with different cybersecurity expenditures and threat profiles face distinct strategic challenges. Our analysis of Hungary, Finland, and the United States illustrates that defensive effectiveness depends not only on absolute spending levels but also on the interaction between budget allocation and country-specific attractiveness to attackers. The second model shifts focus to organizational defense, accounting for heterogeneous attack surfaces through differentiated battlefield values and ratios. Through five organizational case studies spanning critical infrastructure, industrial sectors, service industries, knowledge-based firms, and government entities, we show that optimal defense strategies must be tailored to workforce composition, operational characteristics, and threat actor motivations.

Our findings demonstrate that combining criminological theory with real-world cybercrime data yields actionable insights that transcend generic best-practice recommendations. The models reveal that marginal improvements in defensive capacity or reductions in attacker-attractive characteristics can produce substantive shifts in strategic position, and that even organizations with modest budgets can achieve meaningful protection through targeted interventions. By bridging theoretical frameworks with computational game theory, this work provides a systematic methodology for assessing cyber risk and prioritizing awareness programs based on organizational and national context.

\subsection{Limitations}

While our models provide valuable decision support, several limitations should be noted. First, parameterization relies on cybercrime surveys subject to biases and high-level public statistics that could be too coarse-grained for business grade usage. For organizational models, reliance on open-source intelligence introduces measurement error as many security practices are not publicly disclosed. Nonetheless, data-related limitations can be relaxed when the organization utilizing our models can input its own fine-grained data. Second, we employ simplifying assumptions for tractability: treating attackers as unified strategic actors (abstracting from heterogeneity across opportunistic criminals and state-sponsored APTs), assuming simultaneous rather than dynamic multi-period allocation, modeling deterministic rather than stochastic battlefield outcomes, and presuming uniform marginal costs for defensive interventions across employee roles. 
Third, our battlefield taxonomy reflects current threat intelligence but may not capture emerging attack vectors from generative AI and deepfakes, nor fully represent hybrid organizational forms like platform companies or public-private partnerships. 
Fourth, regret matching approximates Nash equilibria without formal convergence guarantees, and mixed strategies represent average behavior rather than single-instance prescriptions; practical constraints on randomization may limit operational applicability.
Fifth, the model assumes independence in both target selection and battlefields, whereas in practice attackers often return to previously targeted entities, leveraging prior information and partial successes.
Finally, while models identify associations between characteristics and optimal allocations, they do not establish causal mechanisms or capture feedback effects such as reputational deterrence or long-term improvements in employee vigilance.

\subsection{Future work}

Several extensions could improve the model’s theoretical and practical relevance. Dynamic formulations could capture multi-period interactions, adaptive strategies, learning effects, and timing of security investments. Incorporating AI-enabled threats and defenses would reflect how generative AI and deepfakes lower attacker costs and enable hyper-personalized deception, while defensive AI supports anomaly and phishing detection. Heterogeneous attacker models could distinguish opportunistic criminals, organized groups, hacktivists, and state-sponsored actors, allowing threat-specific resource allocation and analysis of adversary competition. Integration with technical defenses could optimize complementary investments in training, access controls, endpoint protection, and network monitoring. Empirical validation through organizational partnerships, field experiments, and longitudinal tracking would test predictive accuracy and intervention effectiveness. Cross-national analyses could identify generalizable patterns, while behavioral extensions would refine understanding of cognitive biases, training efficacy, and cultural influences on security practices. Finally, policy-oriented research could evaluate how regulatory mandates, liability regimes, and disclosure requirements shape incentives, informing strategies that align organizational actions with social welfare.

\subsection*{Acknowledgments}
This work was partially supported by Project
no. 138903 of the Ministry of Innovation and Technology, Hungary, from the NRDI Fund, financed under the FK\_21 funding scheme and a Fulbright Conference Travel Award for Hungarian Alumni.